\documentclass{aa}  

\usepackage{graphicx}
\usepackage{txfonts}
\usepackage[english]{babel}
\usepackage[utf8]{inputenc}
\usepackage{amsmath}
\usepackage{natbib}

\def\asec{\ifmmode ^{\prime\prime}\else$^{\prime\prime}$\fi}

\def\msun{\hbox{M$_{\odot}$}}

\def\degs{\ifmmode ^{\circ}\else$^{\circ}$\fi}
\def\amin{\ifmmode ^{\prime}\else$^{\prime}$\fi}
\def\asec{\ifmmode ^{\prime\prime}\else$^{\prime\prime}$\fi}
\def\farcs{\hbox{$.\!\!^{\prime\prime}$}}  

\def\degs{\ifmmode ^{\circ}\else$^{\circ}$\fi}
\def\amin{\ifmmode ^{\prime}\else$^{\prime}$\fi}

\def\EE#1{\times 10^{#1}}

\def\cm{\mbox{\,cm}}

\def\cm3{\mbox{\,cm$^{-3}$}}
\def\kms{\mbox{\,km~s$^{-1}$}}

\def\kms{\mbox{\,km s$^{-1}$}}

\def\lsim{\!\!\!\phantom{\le}\smash{\buildrel{}\over
 {\lower2.5dd\hbox{$\buildrel{\lower2dd\hbox{$\displaystyle<$}}\over
                                 \sim$}}}\,\,}
\def\gsim{\!\!\!\phantom{\ge}\smash{\buildrel{}\over
{\lower2.5dd\hbox{$\buildrel{\lower2dd\hbox{$\displaystyle>$}}\over
                               \sim$}}}\,\,}
\begin{document}

   \title{No trace of a single-degenerate companion in late spectra of SNe 2011fe and 2014J.}

   \subtitle{}

   \author{P. Lundqvist\inst{1}
          \and
          A. Nyholm\inst{1}
          \and 
          F. Taddia\inst{1}
          \and
           J. Sollerman\inst{1}
           \and
           J. Johansson\inst{2}
           \and
           C. Kozma\inst{1}
           \and
           N. Lundqvist\inst{1}
           \and
           C. Fransson\inst{1}
           \and
           P.M. Garnavich\inst{3}
           \and
           M. Kromer\inst{1}
           \and
           B.J. Shappee\inst{4}
           \and
          A. Goobar\inst{2}
          }

   \institute{Department of Astronomy and The Oskar Klein Centre, AlbaNova University Center, Stockholm University, SE-106 91 Stockholm, Sweden\\
              \email{peter@astro.su.se}
         \and
   Physics Department and The Oskar Klein Centre, AlbaNova University Center, Stockholm University, SE 106 91 Stockholm, Sweden
         \and
   225 Nieuwland Science, University of Notre Dame, Notre Dame, IN 46556-5670, USA
            \and
   Carnegie Observatories, 813 Santa Barbara Street, Pasadena, California 91101, USA 
             }

   \date{Received January 22, 2015; accepted ??, ??}

 
  \abstract
   {}
   {This study aims at constraining the origin of  the nearby Type Ia supernovae (SNe)~2011fe and 2014J. The two most favoured scenarios to 
   trigger the explosion of the white dwarf supernova progenitor is either mass-loss from a non-degenerate companion, or merger 
   with another white dwarf. In the former, there could be a significant amount of left-over material from the companion at the centre 
   of the supernova. Detecting such material would therefore favour the single-degenerate scenario.}
   {The left-over material from a possible non-degenerate companion can reveal itself after about one year, and in this study such
   material has been searched for in spectra of SN~2011fe (at 294 days after the explosion) using the Large Binocular 
   Telescope and for SN~2014J using the Nordic Optical Telescope (315 days past explosion). The observations are interpreted using 
    numerical 
   models simulating the expected line emission from ablated material from the companion star. The spectral lines sought for are 
   H$\alpha$, [O~I]~$\lambda$6300 and [Ca~II]~$\lambda\lambda$7291,7324, and the expected width of these lines 
   is $\sim 1000\kms$, which in the case of the [Ca~II] lines  blend to a broader feature.}
   {No signs of H$\alpha$, [O~I]~$\lambda$6300 or [Ca~II]~$\lambda\lambda$7291,7324 could be traced for any of the two 
   supernovae. When systematic uncertainties are included, the limits on hydrogen-rich ablated gas in SNe~2011fe and 2014J are 
   $0.003~\msun$ and $0.0085~\msun$, respectively, where the limit for SN~2014J is the second lowest ever, and the limit for 
   SN~2011fe is a revision of a previous limit. Limits are also put on helium-rich ablated gas, and here limits from 
   [O~I]~$\lambda$6300 provide the upper mass limits $0.002~\msun$ and $0.005~\msun$ for SNe~2011fe and 2014J, respectively. 
   These numbers are used, in conjunction with other data, to argue that these supernovae can stem from double-degenerate 
   systems, or from single-degenerate systems with a spun up/spun down super-Chandrasekhar white dwarf. For SN~2011fe, other 
   types of hydrogen-rich donors can likely be ruled out, whereas for SN~2014J a main-sequence donor system with large intrinsic 
   separation is still possible. Helium-rich donor systems cannot be ruled out for any of  the two supernovae, but the expected short 
   delay time for such progenitors makes this possibility less likely, especially for SN~2011fe. Published data for SNe~1998bu, 2000cx,
   2001el, 2005am and 2005cf are used to constrain their origin. Finally, the broad lines of SNe~2011fe and 2014J are discussed, and 
   it is found that the [Ni~II]~$\lambda$7378 emission is redshifted by $\sim +1300~\kms$, as opposed to the known blueshift of 
   $\sim -1100~\kms$ for SN~2011fe. [Fe~II]~$\lambda$7155 is also redshifted in SN~2014J.
   SN~2014J belongs 
   to a minority of SNe~Ia that both have a nebular redshift of [Fe~II]~$\lambda$7155 and [Ni~II]~$\lambda$7378, and a slow decline of 
   the Si~II~$\lambda6355$ absorption trough just after $B-$band maximum.}
   {}


   \maketitle
%

\section{Introduction}
It is widely thought that a Type Ia supernova (SN Ia) is the thermonuclear explosion of a carbon/oxygen white
dwarf (WD). The two most common scenarios are that the explosion could be triggered by mass transfer from a non-compact 
companion star \citep[the single-degenerate scenario:][]{whe73,nom82}, or that it is the result of a merger with another 
WD \citep[the double-degenerate scenario:][]{whe73,ibe84,web84}.
While the single-degenerate (SD) scenario where a WD at the Chandrasekhar limit 
accretes matter from a close companion has been the most favoured scenario, there is now growing evidence 
that the double-degenerate (DD) scenario could be the dominant channel for SNe Ia \citep[e.g.,][]{mao14}. 

The lack of knowledge about the true nature of the progenitor systems of SNe Ia is a great disadvantage, since 
they are used as standardisable candles for distance determinations in cosmology \citep[e.g.,][]{goolei11} 
and were used to discover the accelerating expansion of the Universe \citep[e.g.,][]{rie98,per99}. To do 
precision cosmology, systematic effects related to the type of progenitor system should be minimised, 
and possibilities to identify the nature and origin of these systems must be probed. 

One way to constrain the nature of the progenitor systems is to look for merger left-over from the companion star in 
SD scenarios. This could either be done by searching for absorption or emission lines from a circumstellar
medium (CSM) around normal SNe Ia \citep[e.g.,][]{mat05,pat07,sim09,dil12,lun13,ste14}, 
or to identify material blasted off or evaporated from the non-compact companion due to the impact of the SN
ejecta \citep[e.g.,][]{mat05,leo07,sha13b,lun13,mae14}. Of these, absorption lines from a CSM may be the
least conclusive, since such lines may also exist in DD scenarios \citep{shen13}.

Detecting early X-ray or radio emission due to interaction between the 
supernova ejecta and a CSM would argue for a SD scenario, but no such emission has ever been 
observed from a SN Ia \citep[e.g.,][]{panagia06,hug07,han11,rus12}, not even from the very nearby SNe 2011fe 
and 2014J, a fact that has been used to rule out most SD scenarios for those supernovae 
\citep{cho12,mar12, mar14,per14}. There are, however, SD scenarios predicting a very tenuous CSM in
the vicinity of the explosion \citep[e.g.,][]{dis11,jus11,hachisu12}, so non-detections of radio and  X-ray 
emission from SNe Ia are not necessarily fully conclusive in terms of DD or SD scenarios. 

Here we concentrate on probing possible material from a SD companion in late optical spectra of SN 2014J,
the closest SN~Ia for decades. We do this in the same way as was previously done for six other 
SNe~Ia \citep[e.g.,][]{mat05,leo07,lun13,sha13b}, including the nearby SN~2011fe. For the latter, \citet{sha13b} 
claimed an upper limit of $0.001~\msun$ of solar-abundance material to be present in the 
innermost ejecta of the supernova, which is in clear conflict with hydrodynamical simulations of SD 
scenarios \citep{mar00,pak08,liu12,liu13a,pan12}. For the other five SNe~Ia, the upper mass limit was 
$0.01-0.03~\msun$, which is in marginal conflict with SD scenarios. 

A way to avoid conflict between the lack of hydrogen lines in late spectra and hydrodynamical models is to assume that the 
SD companion was helium-rich. In such a case, $0.0024-0.028~\msun$ \citep{pan12,liu13b} of helium-rich material may 
instead pollute the innermost ejecta of the SN~Ia. In \citet{liu13b} it was suggested to look for helium lines in this situation, 
but as discussed in \citet{lun13}, helium lines, due to this pollution, are not expected to be as prominent  as lines of oxygen and 
calcium. So, in addition to look for hydrogen via H$\alpha$, we will here also search for oxygen and calcium 
lines with a width of $\sim 1000\kms$. 

In our analysis, we use the same computer code to calculate the line emission from ablated mass from the 
SD companion as in our previous similar analyses \citep{mat05,lun13}, i.e., the model discussed
in \citet{lun13}, which is based on calculations for the W7 model \citep{nom84,thi86}. Details of the modeling
of late SN~Ia spectra using this code for W7 are described in \citet{sol04}, and in \citet{koz05} for other 
explosion models. The results in \citet{mat05} for the modelled H$\alpha$ luminosity were extrapolated
by \citet{leo07} and \citet{sha13b} to obtain the limits on ablated mass in those studies.

While SN~2014J is much closer to us than SN~2011fe, SN~2014J is more extinguished. One could for SN~2014J therefore
expect to be about as sensitive in terms of limits on polluting mass from a SD donor, as was reported for SN~2011fe \citep{sha13b}. 
For a comparison between the two
supernovae, we include them both in our analysis. Throughout the paper we adopt the distances 6.1 Mpc 
and 3.4 Mpc to SNe~2011fe and 2014J, respectively. For the Galactic extinction ($R_V = 3.1$) we use 
$E(B-V) = 0.026$~mag for SN~2011fe and  $E(B-V) = 0.06$~mag for SN~2014J. For SN~2014J we also add 
$E(B-V) = 1.37$~mag
($R_V = 1.4$) for M82. We refer to \citet{ama14}, \citet{fol14}, \citet{goo14a} and \citet{joh14} for a discussion 
on those values. For the recession velocities to the supernovae we use the host galaxy recessions, 
i.e., $+241\pm2 \kms$ for SN~2011fe and $+203\pm4 \kms$ for SN~2014J \citep{dev91}. 
In Section 2 we describe our observations and the data, in Section 3 we show our results, and in Section 4 we 
provide a discussion. Finally, in Section 5 we make our conclusions.

\begin{figure*}
\centering
\includegraphics[width=\hsize,clip=]{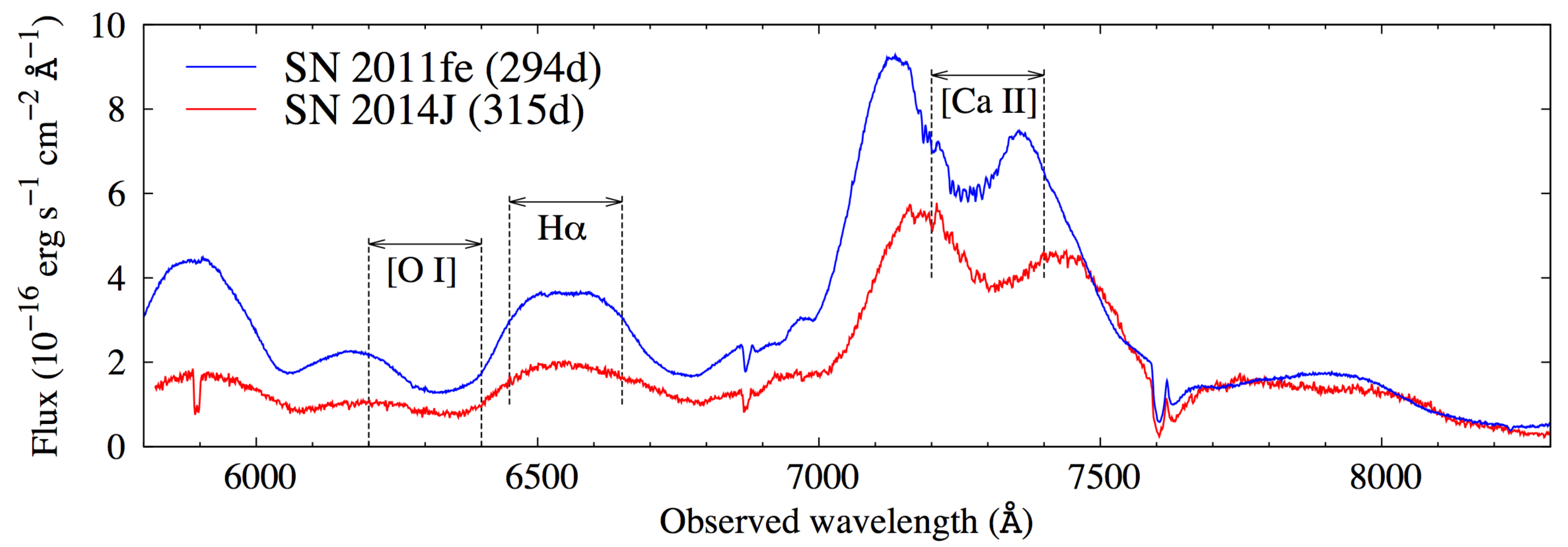}
\caption{Observed LBT spectrum (blue colour) of SN 2011fe at 294 days after the explosion 
\citep[cf.][for the full LBT spectrum of the supernova]{sha13b} and observed NOT spectrum 
(red colour) of SN~2014J at 315 days after the explosion. No reddening or redshift correction has been made to the spectra.
The recession velocities of the SNe are similar, and only 241$\kms$ and 203$\kms$, respectively.
The wavelength regions of interest for this study are marked with dashed lines and arrows. Those wavelength regions
are the same as in Figs. \ref{fig:11fe_spectra} and \ref{fig:14J_spectra}. Note the general blueshift of the broad double-peak
between $7000-7500$~\AA\ for SN~2011fe relative to SN~2014J, whereas for the peak between $6400-6750$~\AA, there
is no obvious shift in wavelength. See text for further details. 
              }
\label{fig:14J_spectrum}
\end{figure*}

\section{Observations}
\subsection{Observations of SN~2011fe}
SN 2011fe was observed on 2012 Jun 12.16 (JD 2456090.66), i.e., 294 days after the explosion on 
August 23.7, 2011, for 2 full hours with the MODS spectrograph on the  8.4-meter LBT\footnote{The Large Binocular 
Telescope (LBT) and Multi-Object Dual Spectrograph \citep[MODS;][]{pog13}.}. The spectrum was first 
published by \citet{sha13b}, and then later also by \citet{mcc13}. We have absolute calibrated the spectrum by comparison 
to the $R-$band photometry by \citet{mun13}. The spectrum is shown in Fig. \ref{fig:14J_spectrum}.

\subsection{Observations of SN~2014J}
We observed SN~2014J with the Nordic Optical Telescope (NOT) on November 26, 2014,  i.e., 315 days 
after the explosion on January 14.75, 2014 \citep{zheng14}. The NOT observations were made in service mode under NOT proposal 50-023 (P.I. A. Nyholm). The ALFOSC spectrograph was used with grism \#8 and a $1\farcs0$ slit (parallactic slit orientation) to get four longslit spectra with exposure time 1800 seconds per spectrum. The setup allowed 
us to obtain spectral coverage of the interval  $5820 - 8370$ \AA, where the features of interest in this 
investigation can be found. The expected resolution for our setup with a $1\farcs0$ slit is 
$\Delta \lambda = 7.0$ \AA. The resolution of the obtained spectra was estimated using night sky lines on each side of H$\alpha$ in the individual spectra, and was found to be between 7.1 \AA\ and 7.8 \AA. At the time of the observations, there were thin clouds and variable seeing (extremes: $0\farcs9$ and $1\farcs3$). UT date for the mid-exposure time of the first and last frames was 2014 Nov 26.11 and 2014 Nov 26.17. The spectra were obtained in the airmass range 1.43 -- 1.74, and were reduced in the standard way with IRAF\footnote{IRAF is distributed by the National Optical Astronomy Observatories, which are operated by the Association of Universities for Research in Astronomy, Inc., under cooperative agreement 
with the National Science Foundation.} applying overscan corrections, bias subtractions and flat fields to 
the individual spectra. The flat fields used were taken with the
telescope pointing at the SN, directly before the SN spectra themselves
were  taken. A He-Ne lamp was used for the wavelength calibration and Feige 34 was used as 
flux standard for the four extracted, sky-subtracted SN spectra. The four individual SN spectra were then
co-added. The $R-$band (Bessel) magnitude $17.599\pm0.041$ of SN~2014J had been measured with the NOT on November 25, 2014 and this photometry was used for the 
absolute flux calibration of the final co-added spectrum. 
The co-added and flux-calibrated 
spectrum is shown in Fig. \ref{fig:14J_spectrum}.
%
%
%

\begin{figure*}
\setlength{\unitlength}{1mm}
\resizebox{19cm}{!}{
\begin{picture}(190,240)(0,0)
\put (0,175) {\includegraphics[width=99mm, angle=0,clip=]{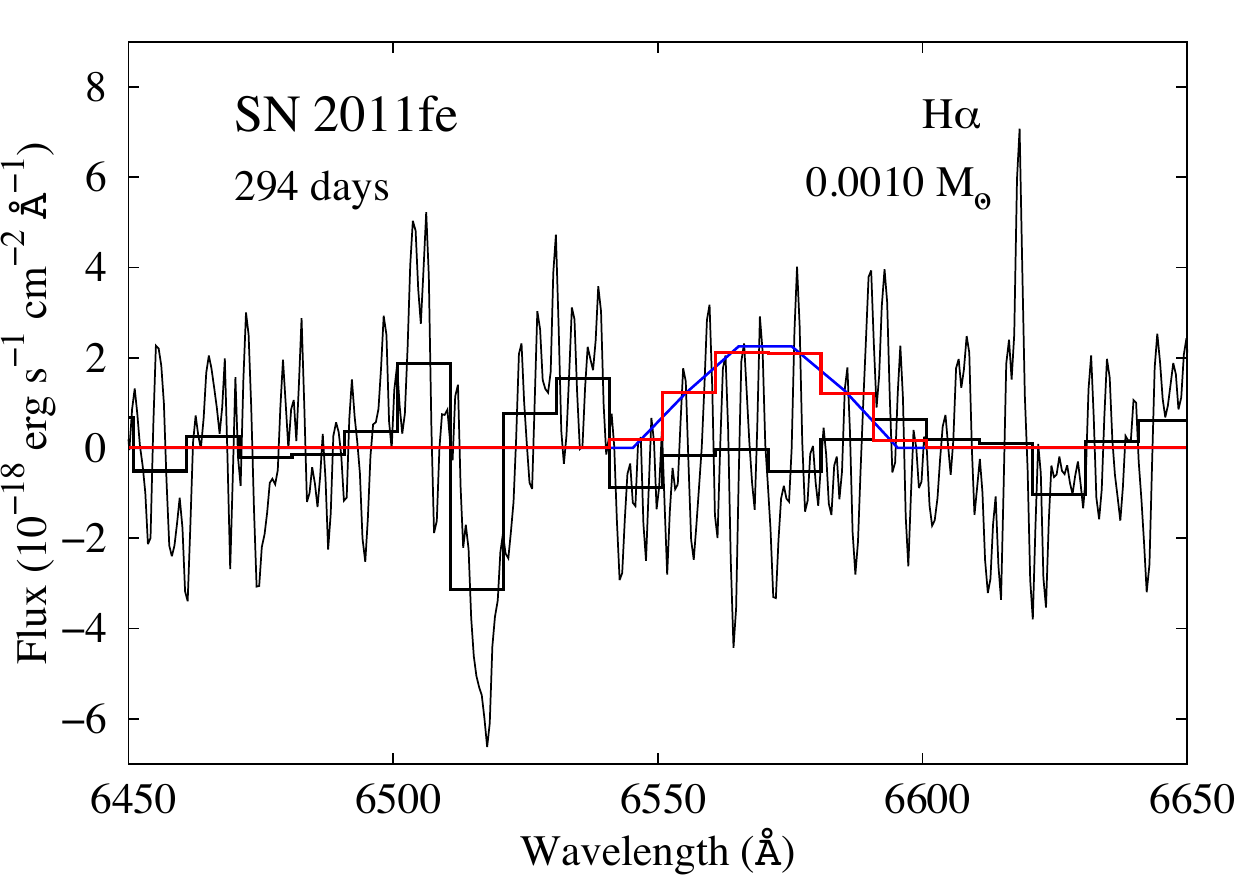}} 
\put (98,175) {\includegraphics[width=88mm, angle=0,clip=]{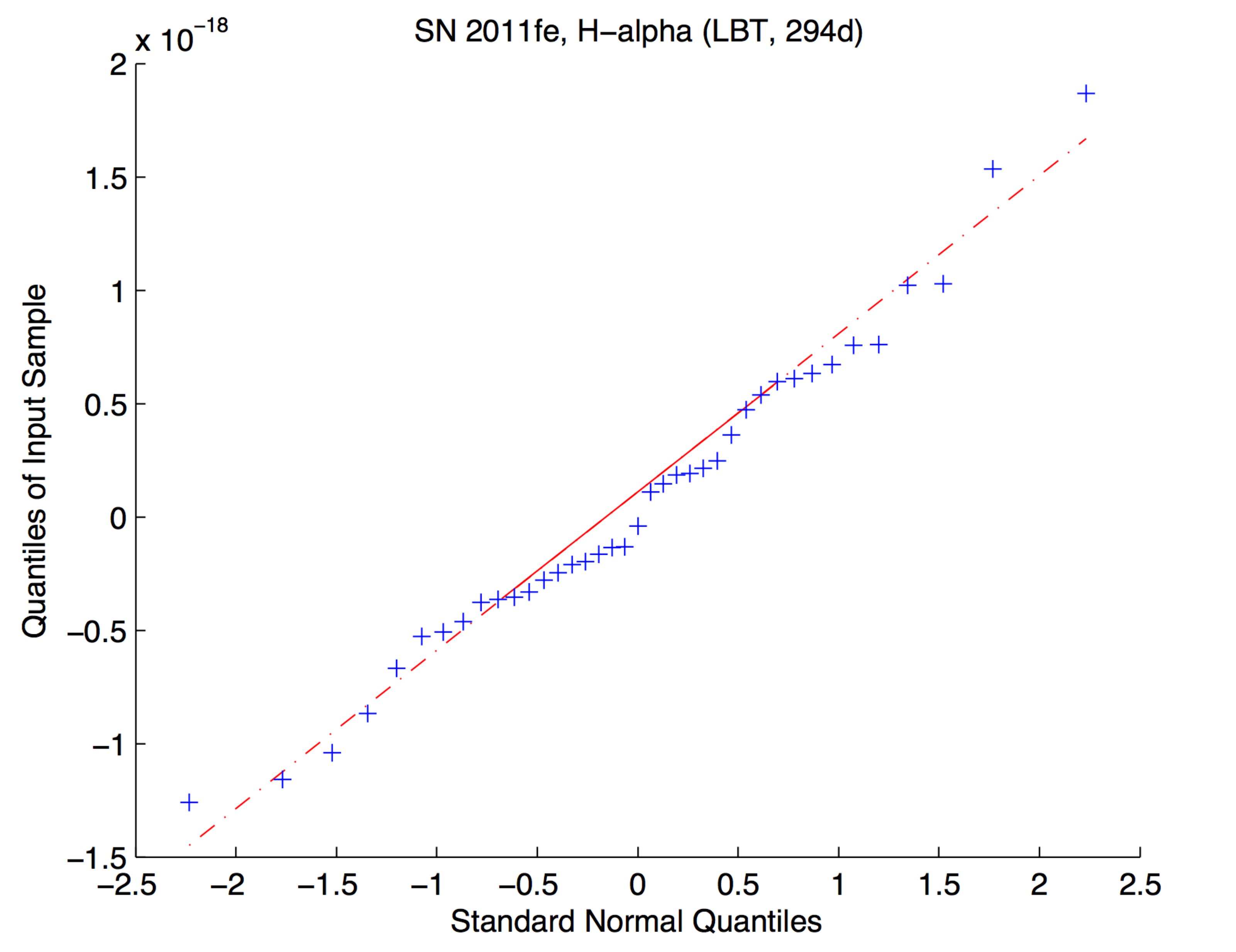}} 
\put (0,103) {\includegraphics[width=99mm, angle=0,clip=]{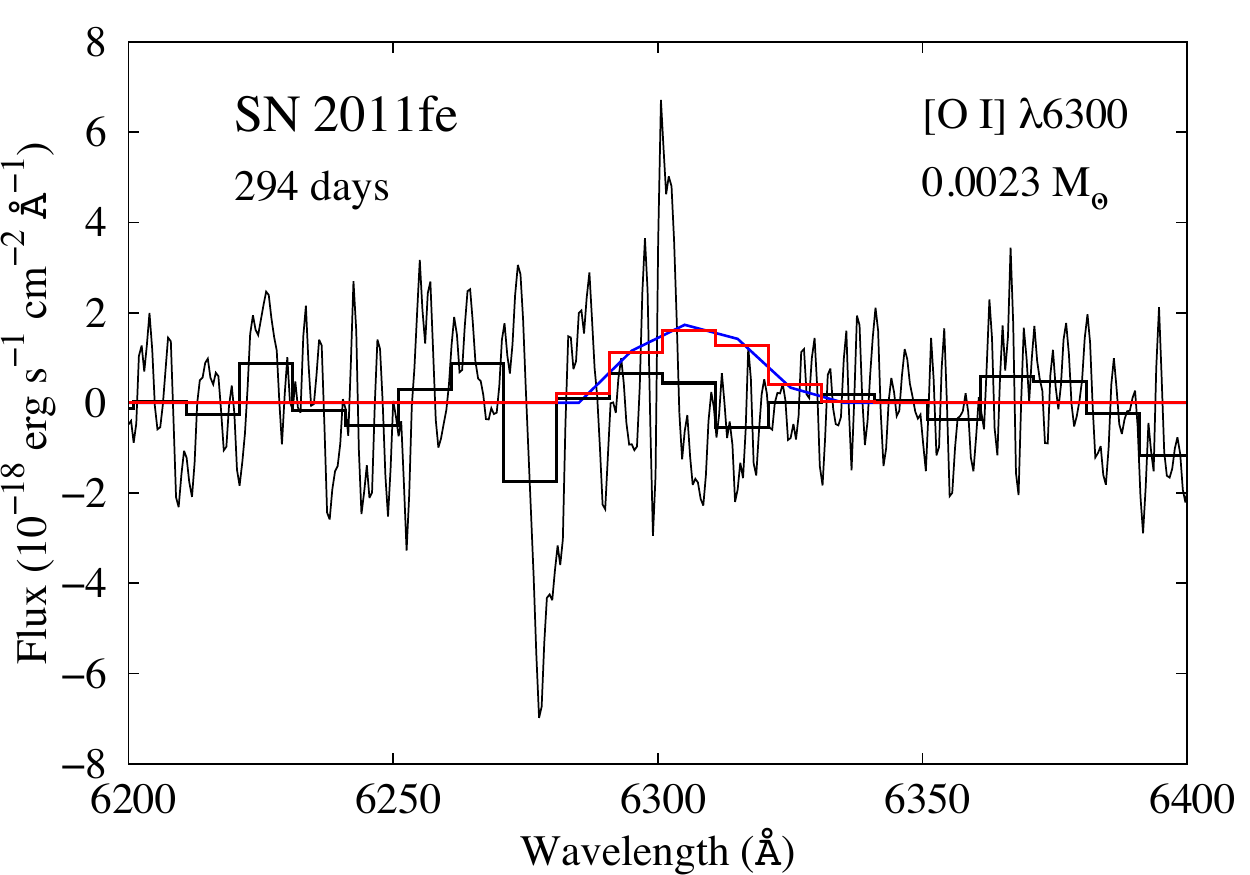}} 
\put (98,103) {\includegraphics[width=88mm, angle=0,clip=]{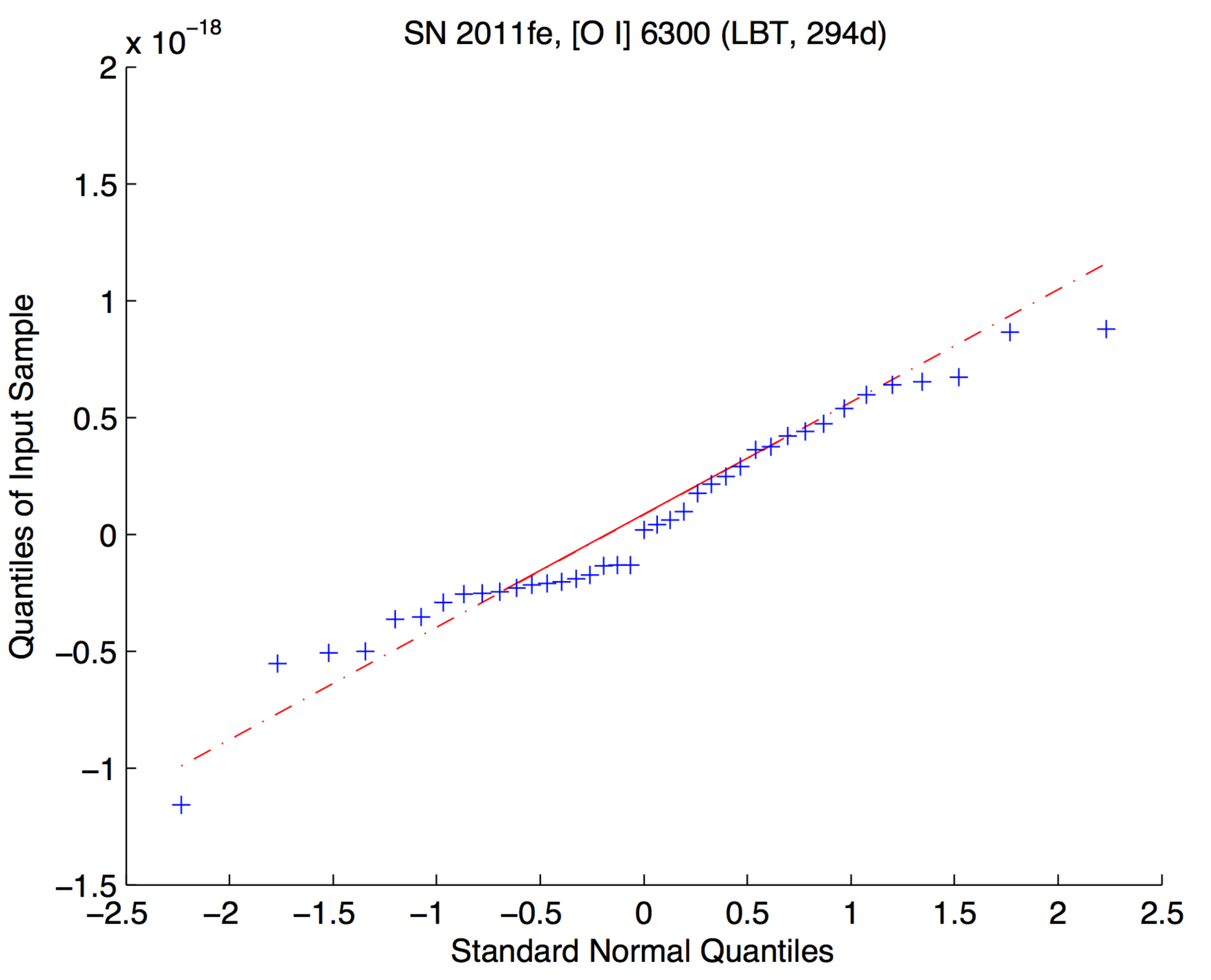}} 
\put (0,31) {\includegraphics[width=99mm, angle=0,clip=]{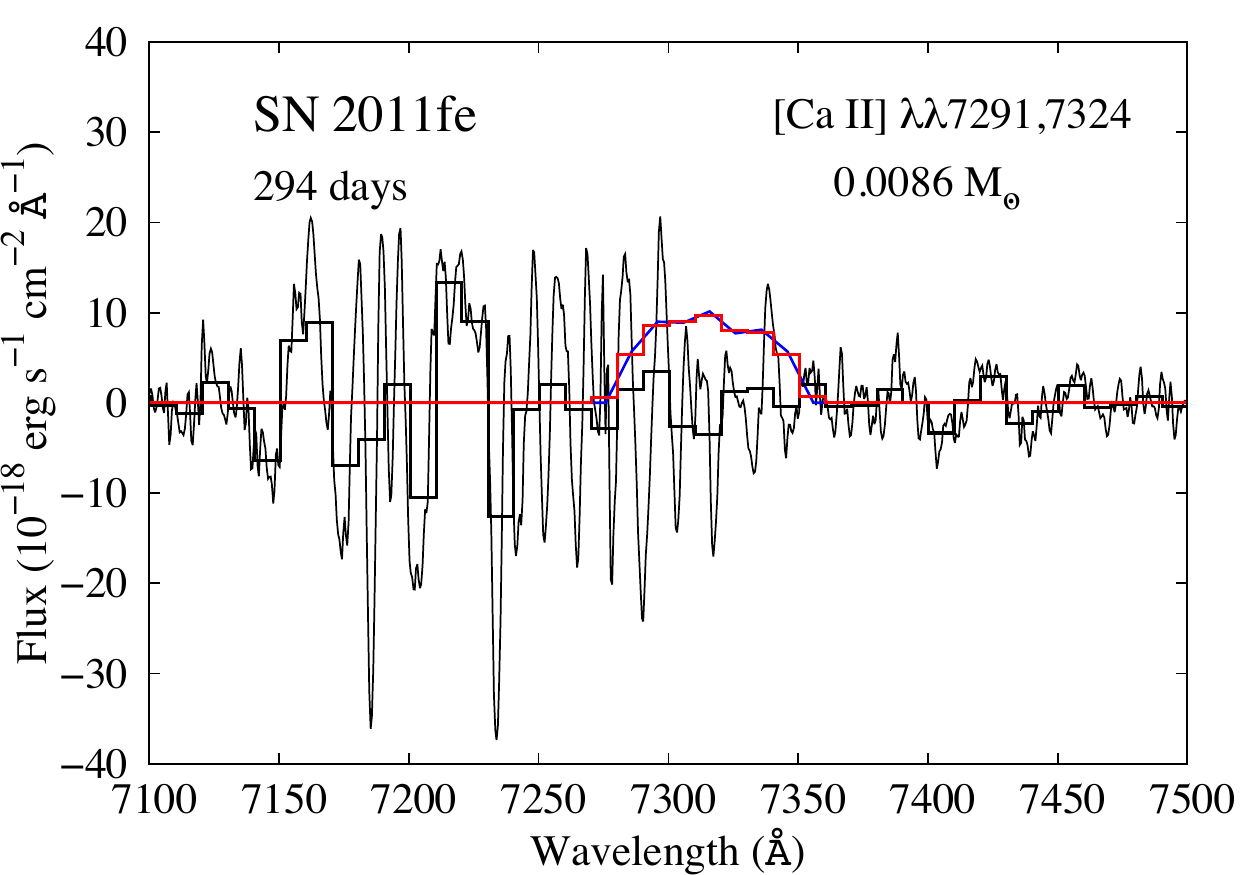}} 
\put (98,31) {\includegraphics[width=88mm, angle=0,clip=]{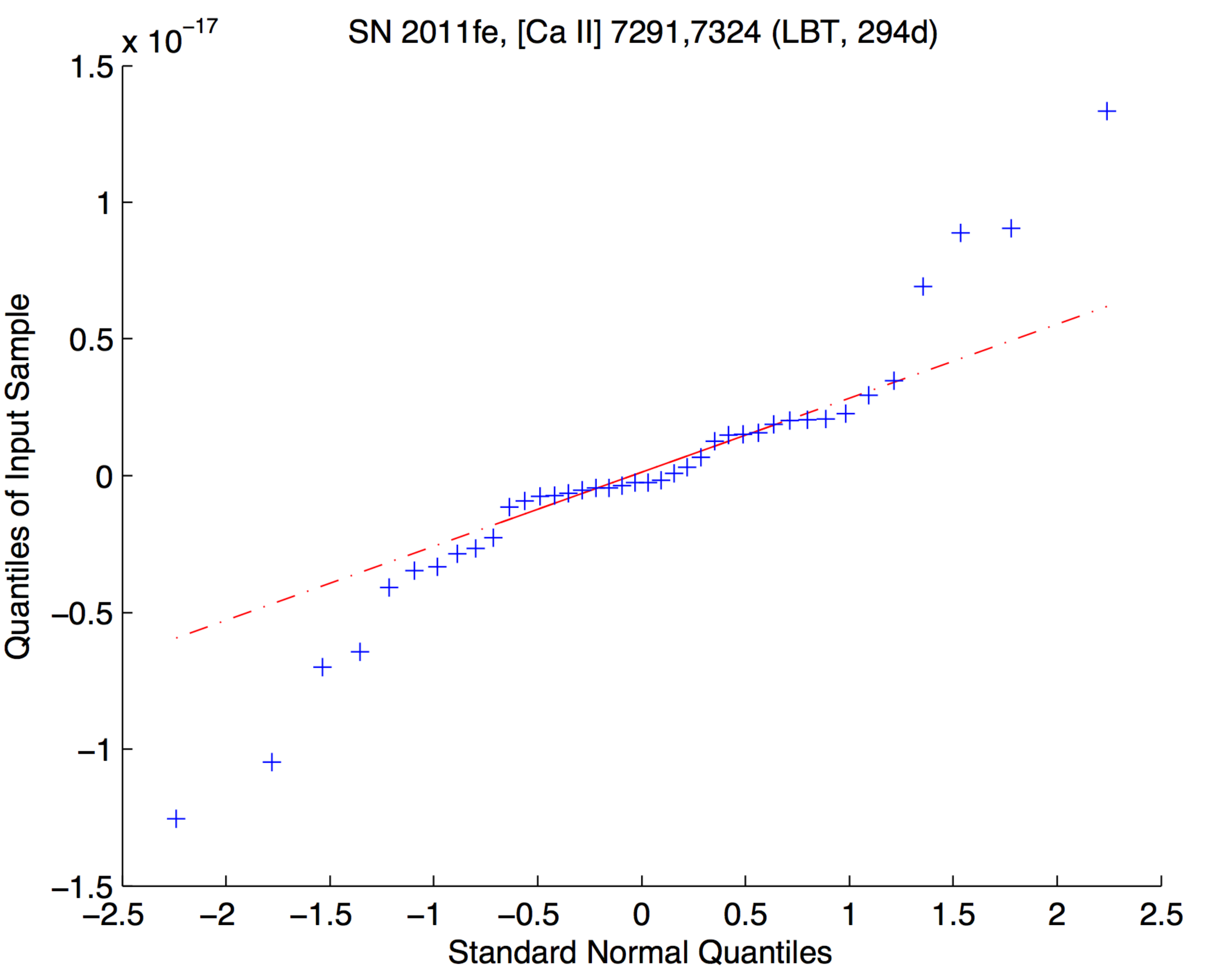}} 
\put (0,32) {\parbox[t]{185mm}{\caption{
{\it Left columns:} LBT net spectra (i.e., spectra after continuum removal) of SN~2011fe 294 days after the 
explosion (thin black lines), in the spectral regions around H$\alpha$ (upper panel), [O~I]~$\lambda6300$ (middle
panel) and [Ca~II]~$\lambda\lambda$7291,7324 (lower panel), respectively. The thick black histogram 
lines show the observed spectrum after 10~\AA\ binning. No correction for redshift was made. The blue lines show the modelled 
line emission, using the model in \citet{lun13} for 294 days. The red histogram lines show the modelled flux binned to the 
same resolution as the binned observed spectrum. The mass of solar-metallicity material in these models are
$0.0010~\msun$, $0.0023~\msun$ and $0.0086~\msun$, respectively, and correspond to estimated $3\sigma$ statistical
upper limits of the mass. The modelled spectra have been redshifted by $+203~\kms$ and reddened by $E(B-V) =0.026$~mag 
to match the velocity and extinction of the supernova. A distance of 6.1 Mpc was used. The mass limit for
H$\alpha$ agrees with that of \citet{sha13b} using the same data. {\it Right columns:} Quantile-quantile plots of the data in 
the net spectra. Blue data points are from the distribution of binned fluxes (in 10~\AA\ bins) for 400~\AA\ spectral regions
around the modelled lines. The dashed red lines are for the simulated normal distribution. As can be seen, the data samples do
not deviate appreciably from a normal distribution, except for the  [Ca~II] lines. The spectral bins dominated by telluric features 
at $\sim 6275$~\AA\ and $\sim 6520$~\AA\  \citep[cf.][]{sha13b} were removed from the sample prior to analysis. See text for 
further details.
}\label{fig:11fe_spectra}}}
\end{picture}}
\end{figure*}

\section{Results}
In Sect. 3.1 \& 3.2 we discuss a strict statistical approach to estimate upper limits on the tentative line emission from the ablated gas.
In Sect. 3.3 we evaluate to what extent these statistical limits can be used to really set lower limits, or whether systematic effects 
dominate.

\subsection{SN~2011fe}
With the LBT spectrum we initiated our statistical approach in a similar way to what was done in \citet{sha13b}, i.e., we smoothed 
the spectrum using a second-order Savitzky-Golay polynomial \citep{pre92}. We then subtracted the non-smoothed spectrum 
from the smoothed one, creating a net spectrum. Like \citet{sha13b}, we found that a smoothing width 
of $\pm 30$~\AA\ produced optimal net spectra. 

Net spectra are shown as thin black solid lines in the left panels of Fig. \ref{fig:11fe_spectra} for the spectral regions 
around H$\alpha$, [O~I]~$\lambda6300$ and [Ca~II]~$\lambda\lambda$7291,7324, i.e., lines that are expected from ablated 
gas of a SD companion \citep{lun13}. For our further analysis, we binned the net spectra in 10~\AA\ bins, which is also shown in 
these figure panels as thick black solid lines. The 10~\AA\ binning was chosen to obtain enough number of spectral bins per
expected line and to get enough number of spectral bins to study the noise within a reasonable wavelength region (see below). At the
same time, 10~\AA\ binning is fine enough to detect narrow absorption and emission features not arising in the SN.

To investigate the noise distribution of the binned net spectra in the wavelength regions of H$\alpha$, 
[O~I]~$\lambda6300$ and [Ca~II]~$\lambda\lambda$ 7291,7324, we sampled fluxes in 40 wavelength 
bins around the wavelength of the modelled spectral lines, and compared that to the normal distribution, 
using a quantile-quantile test \citep{rice07}. Prior to estimating the standard deviations we removed the spectral 
bins including the features at $\sim 6275$~\AA\ and $\sim 6520$~\AA, marked as telluric features by \citet{sha13b}. 
In a quantile-quantile plot, a deviation from a straight line reveals a non-Gaussian distribution.
As can be seen in the right panels of Fig. \ref{fig:11fe_spectra}, the noise does not deviate appreciably from 
that of a normal distribution, except for the [Ca~II] lines. 

The estimated standard deviation, and its estimated 
95\% confidence level \citep{rice07}, for the
$6350 - 6750$~\AA, $6100 - 6500$~\AA\ and $7100 - 7500$~\AA\ spectral regions are 
$(6.90^{+2.10}_{-1.19})\EE{-19}$ erg~s$^{-1}$~cm$^{-2}$~\AA$^{-1}$,
$(4.43^{+1.35}_{-0.76})\EE{-19}$ erg~s$^{-1}$~cm$^{-2}$~\AA$^{-1}$ and
$(4.70^{+1.41}_{-0.80})\EE{-18}$~erg~s$^{-1}$~cm$^{-2}$~\AA$^{-1}$
for each spectral region, respectively. For each spectral region, we used the maximum standard deviation within its 95\% confidence 
level range to make a robust estimate of the 1$\sigma$ noise of the spectral bins of the expected line profiles.  

\begin{figure*}
\setlength{\unitlength}{1mm}
\resizebox{19cm}{!}{
\begin{picture}(190,240)(0,0)
\put (0,175) {\includegraphics[width=99mm, angle=0,clip=]{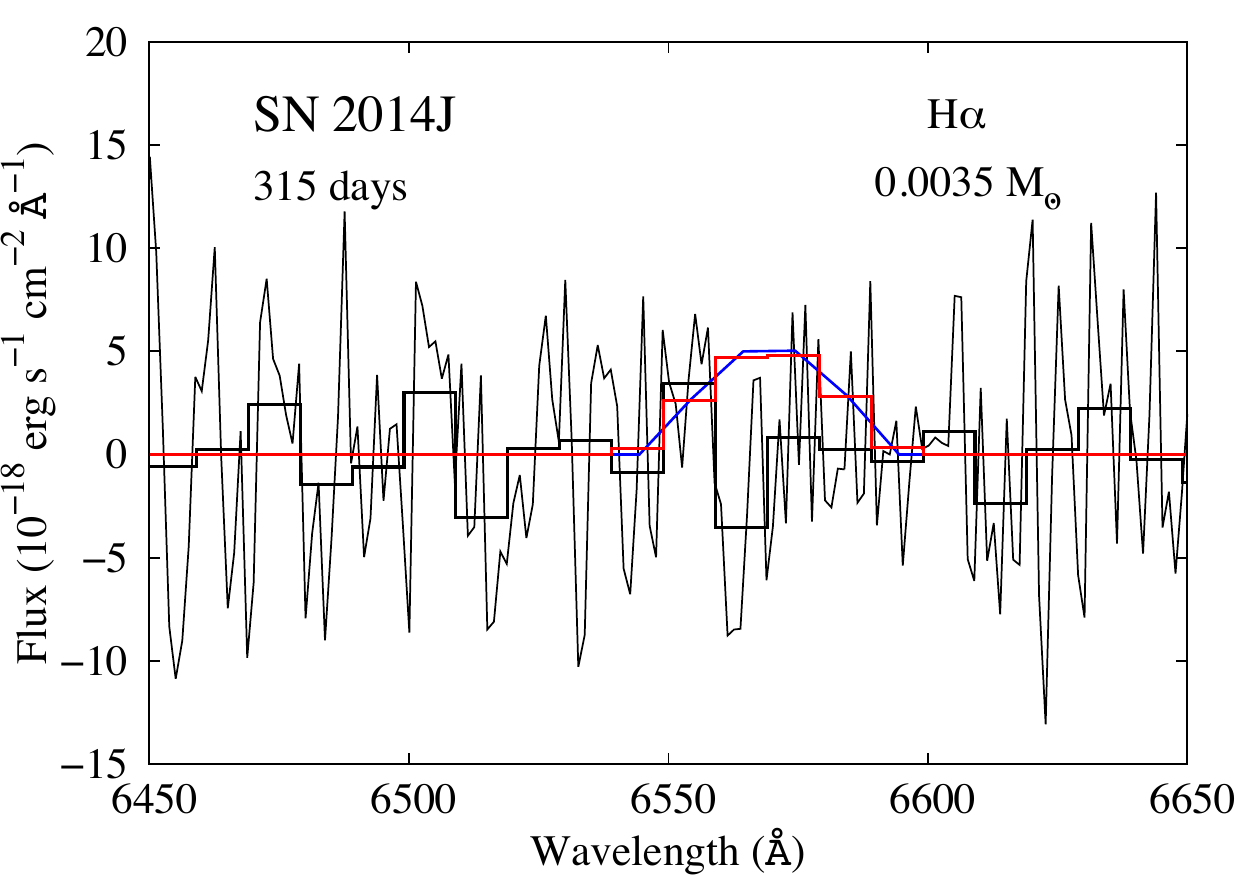}} 
\put (98,175) {\includegraphics[width=88mm, angle=0,clip=]{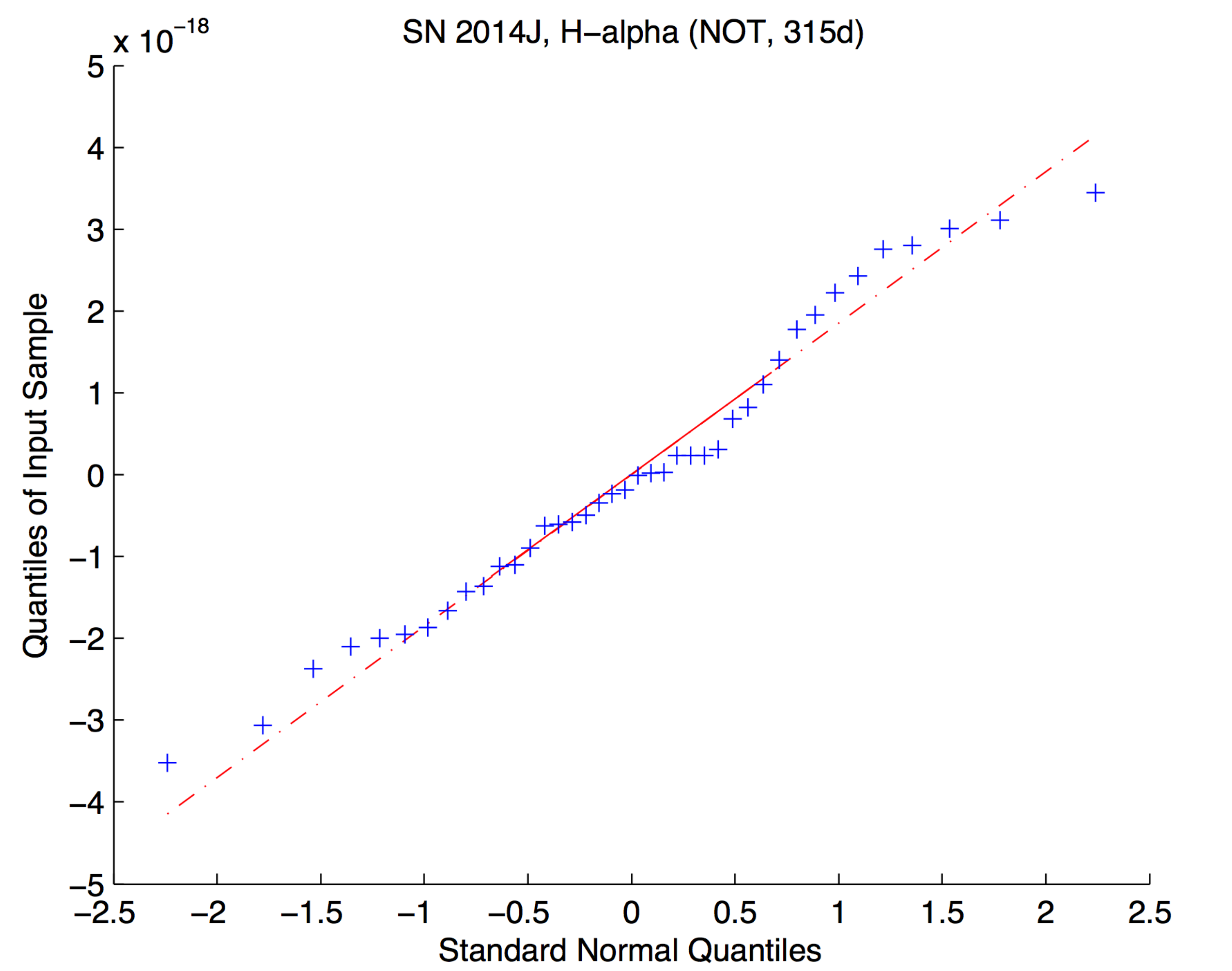}} 
\put (0,103) {\includegraphics[width=99mm, angle=0,clip=]{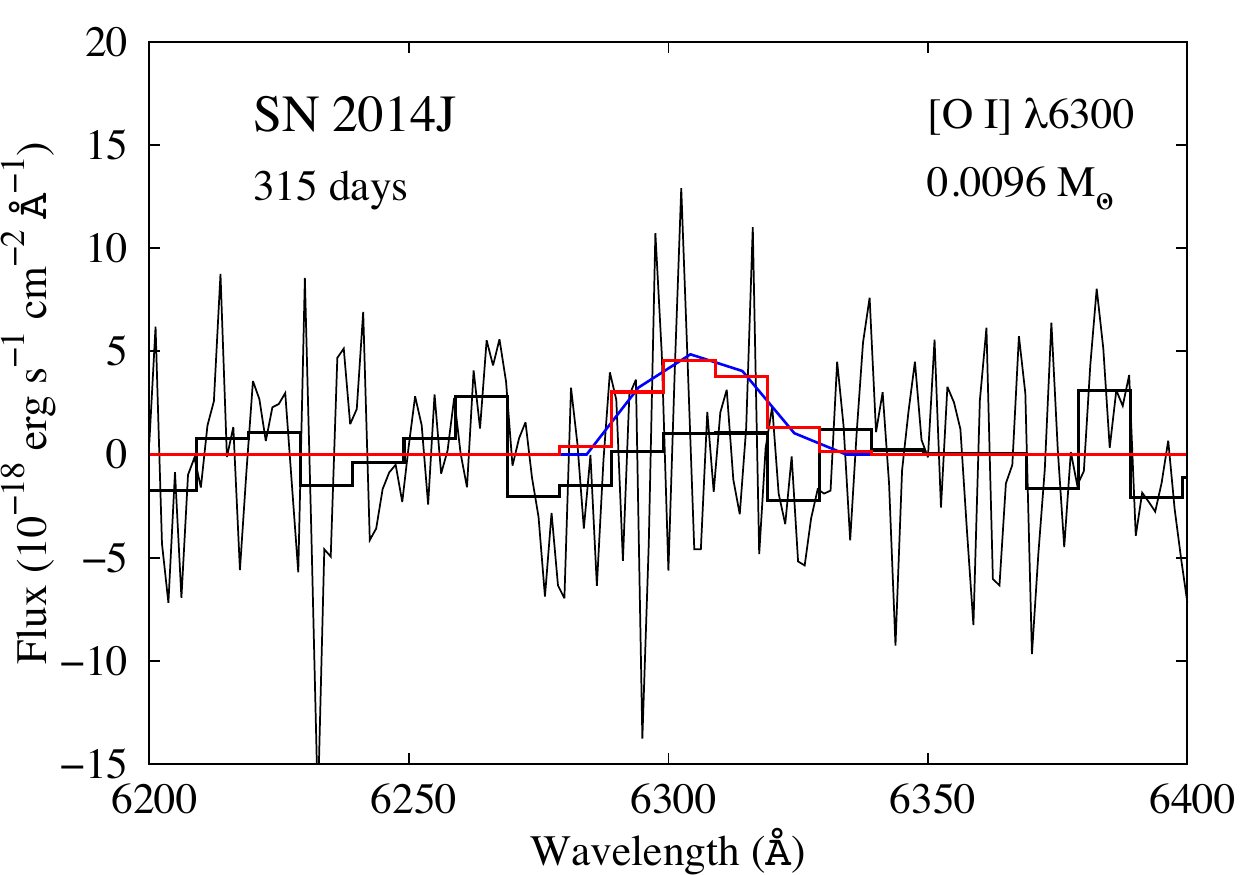}} 
\put (98,103) {\includegraphics[width=88mm, angle=0,clip=]{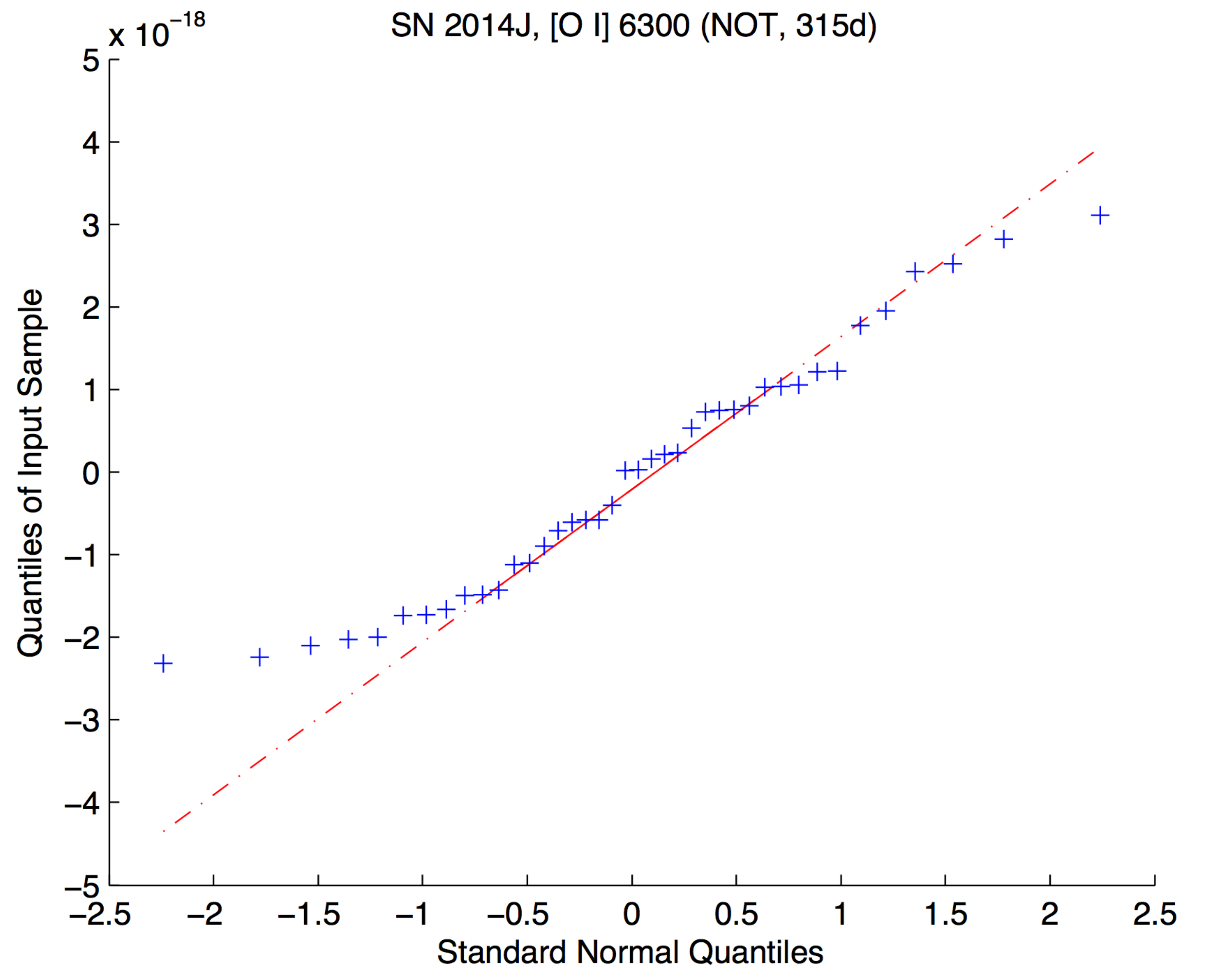}} 
\put (0,31) {\includegraphics[width=99mm, angle=0,clip=]{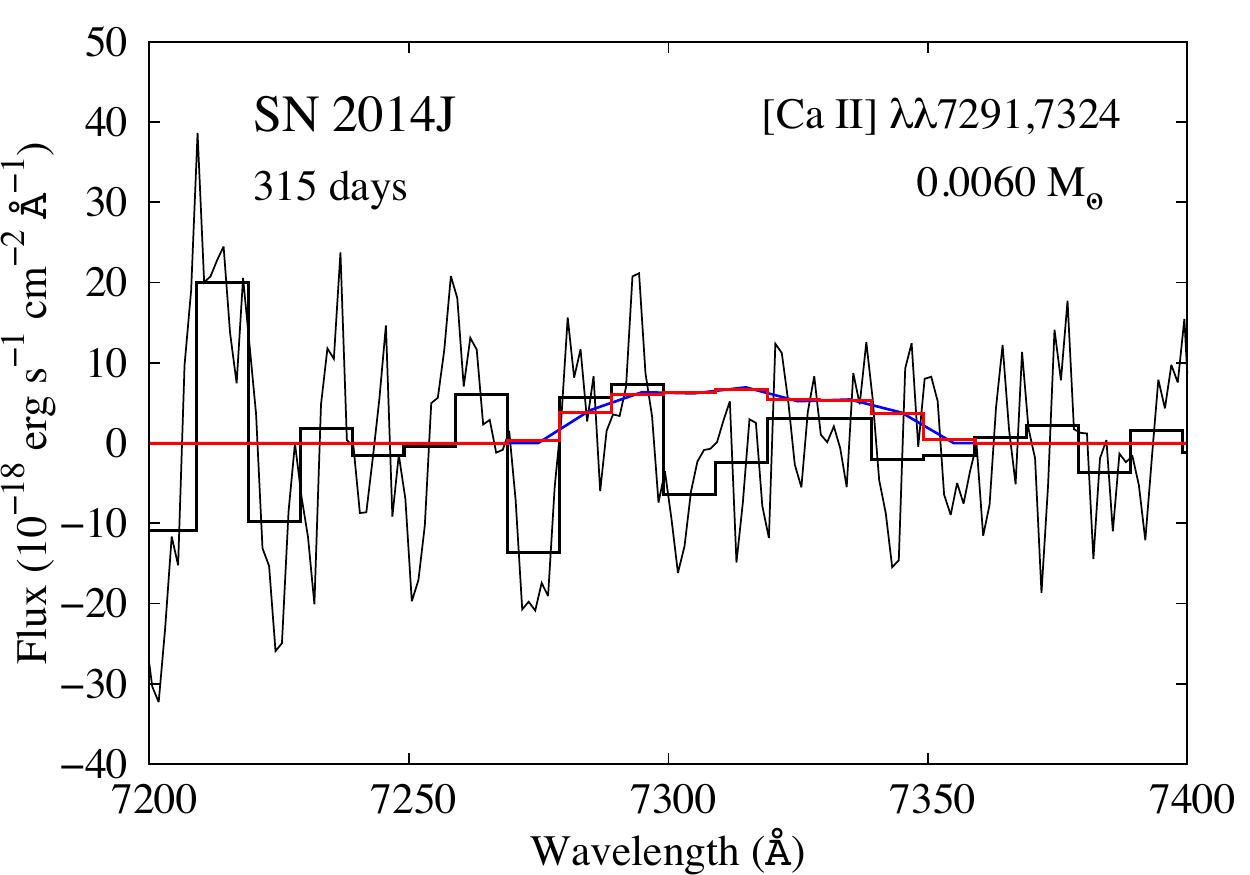}} 
\put (98,31) {\includegraphics[width=88mm, angle=0,clip=]{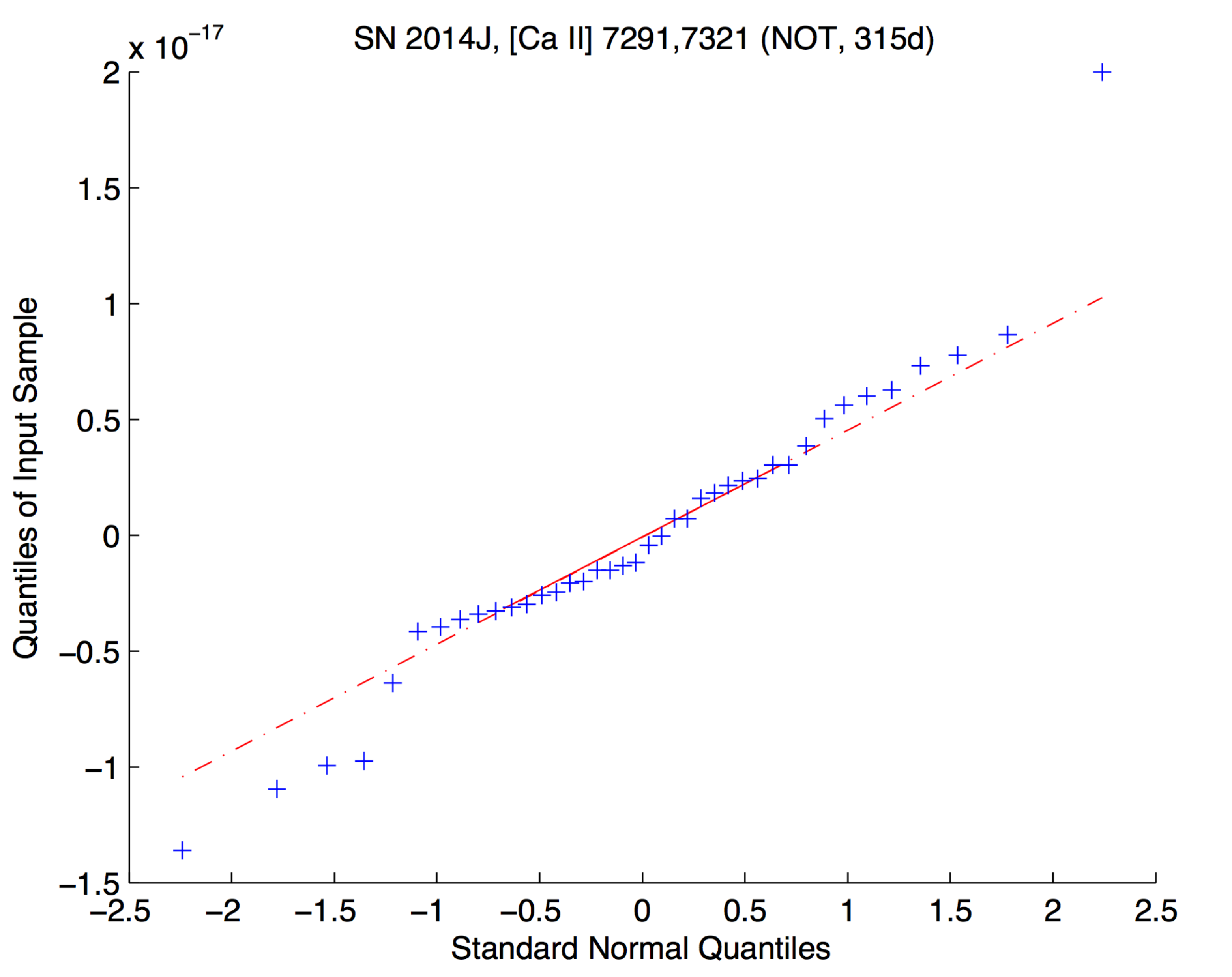}} 
\put (0,32) {\parbox[t]{185mm}{\caption{
{\it Left columns:} NOT net spectra (i.e., spectra after continuum removal) of SN~2014J around H$\alpha$ (upper panel), 
[O~I]~$\lambda6300$ (middle panel) and [Ca~II]~$\lambda\lambda$7291,7324 (lower panel), respectively. The thick black 
histogram lines show the observed spectrum after 10~\AA\ binning. No correction for redshift was made. The blue lines show the 
modelled line emission, using the model in \citet{lun13} for 315 days. The red histogram lines show the modelled flux binned to the 
same resolution as the binned observed spectrum. The mass of solar-metallicity material in these models are
$0.0035~\msun$, $0.0096~\msun$ and $0.0060~\msun$, respectively, and correspond to estimated $3\sigma$ statistical upper limits
of the mass.  The modelled spectrum has been redshifted by $+241~\kms$ and reddened according to what is given in Section 1 
to match the redshift and extinction of the supernova. A distance of 3.4 Mpc was used. {\it Right columns:} Quantile-quantile plots of 
the data in the net spectra. Blue data points are from the distribution of binned fluxes (in 10~\AA\ bins) for 400~\AA\ spectral regions
around the modelled lines. The dashed red lines are for the simulated normal distribution. As can be seen, the data samples do
not deviate appreciably from a normal distribution, except for the strongest absorption features around the [O~I] line, and the strongest
emission feature in the [Ca~II] spectrum in the lower left panel. No spectral bins were, however, removed from the sample prior to 
analysis. See text for further details.
}\label{fig:14J_spectra}}}
\end{picture}}
\end{figure*}

Model spectra were obtained from interpolation in time, and inter- and extrapolation in ablated mass using the grid of models in
\citet{lun13} where the masses were varied between $0.01-0.50~\msun$. Extrapolation to lower masses than $0.01~\msun$ works well 
for all lines considered here as collisional deexcitation of even [O~I]~$\lambda$6300 is unimportant for such low masses.
We applied appropriate redshift and reddening to the model spectra and mapped them onto the 10~\AA\ spectral 
grid and created 10\,000 artificial spectra by adding noise using the Monte Carlo method, where the noise was estimated from the
maximum standard deviations at 95\% confidence level. We ranked the 
simulated line fluxes, and for each line (or doublet in case of [Ca~II]), we estimated 1$\sigma$ errors from those 
ranked in places 1587 and 8413. A 3$\sigma$ statistical upper limit to the ablated mass was estimated from the
mass that gives a 1$\sigma$ limit of the flux that is three times lower than the modelled line flux. The left panels of
Fig. \ref{fig:11fe_spectra} show the modelled line profiles (in blue) for those masses, as well as when mapped 
onto the spectral grid (red histogram lines). Note that for [O~I] we excluded the weak [O~I]~$\lambda6364$ component
from the analysis, as this does not add any important constraints on the oxygen mass. 



For H$\alpha$ we estimate that solar-metallicity ablated material with a mass of $0.0010~\msun$ would 
have been enough to detect in the LBT spectrum. This is fully consistent with the upper limit of $0.001~\msun$ 
reported by \citet{sha13b}. For [O~I]~$\lambda6300$ and [Ca~II]~$\lambda\lambda$7291,7324 we estimate 
that at least $0.0023~\msun$ and  $0.0086~\msun$, respectively, of ablated material with solar-metallicity is 
needed to result in a detection. These masses correspond to observed line fluxes of $7.0\EE{-17}$~erg~s$^{-1}$~cm$^{-2}$,
$4.6\EE{-17}$~erg~s$^{-1}$~cm$^{-2}$ and $5.5\EE{-16}$~erg~s$^{-1}$~cm$^{-2}$, respectively. The relatively
high limit from the [Ca~II] lines is due to a noisy region in the spectrum between $\sim 7150-7350$~\AA\ (cf. Fig. 1) hampered
by the atmosphere.

\subsection{SN~2014J}
 
The NOT spectrum of SN~2014J was also smoothed using a second-order Savitzky-Golay polynomial, and from this a net 
spectrum was created, similar to what was done for SN~2011fe in Sect. 3.1. Again, we smoothed the spectrum with a width 
of $\pm 30$~\AA. The net spectra for the spectral regions around H$\alpha$, [O~I]~$\lambda6300$ and 
[Ca~II]~$\lambda\lambda$7291,7324 are shown as thin black solid lines in the left panels of Fig. \ref{fig:14J_spectra}. 
For the further analysis, we binned the net  spectra in 10~\AA\ bins, shown in these panels of Fig. \ref{fig:14J_spectra}
as thick black solid lines.

The noise distribution of the binned net spectra in the wavelength regions of H$\alpha$, 
[O~I]~$\lambda6300$ and [Ca~II]~$\lambda\lambda$7291,7324, was investigated in the same way as for SN~2011fe in
Sect. 3.1. The quantile-quantile plots for the noise distribution around those lines are shown in the right panels of 
Fig. \ref{fig:14J_spectra}. Unlike for SN~2011fe, we did not remove any outliers prior to the noise estimate.  
As can be seen in the right panels of Fig. \ref{fig:14J_spectra}, the noise is represented well by that
of a normal distribution. There is only one obvious outlier, and that is the emission feature around $7210$~\AA\ in the
lower left panel of Fig. \ref{fig:14J_spectra} (cf. Sect. 4.3). 

The estimated standard deviation, and its 
estimated 95\% confidence level, for the
$6350 - 6750$~\AA, $6100 - 6500$~\AA\ and $7100 - 7500$~\AA\ spectral regions are 
$(1.78^{+0.53}_{-0.30})\EE{-18}$ erg~s$^{-1}$~cm$^{-2}$~\AA$^{-1}$,
$(1.51^{+0.45}_{-0.26})\EE{-18}$ erg~s$^{-1}$~cm$^{-2}$~\AA$^{-1}$ and
$(6.06^{+1.82}_{-1.03})\EE{-18}$~erg~s$^{-1}$~cm$^{-2}$~\AA$^{-1}$
for each spectral region, respectively. As for SN~2011fe in Sect. 3.1, we used the maximum standard deviation 
within its 95\% confidence level range to estimate the 1$\sigma$ statistical uncertainty for the spectral bins of the 
expected line profiles. Comparing the standard deviation uncertainties for SN~2014J with the 1$\sigma$ uncertainties 
for SN~2011fe, it can be noted that the NOT spectra are somewhat noisier than the LBT spectra in the spectral regions 
of H$\alpha$ and [O~I]~$\lambda6300$, whereas the noise levels are about the same for the 
[Ca~II]~$\lambda\lambda$7291,7324 spectral region. This agrees with a visual inspection of the left panels of 
Figs. \ref{fig:11fe_spectra} and \ref{fig:14J_spectra}.

For H$\alpha$ and  [O~I]~$\lambda6300$ we estimate that solar-metallicity ablated material with a mass of $0.0035~\msun$ 
and $0.0096~\msun$, respectively, would have been enough to detect these lines in the NOT net spectrum. These are factors of $3-5$ 
higher than for SN~2011fe, mainly due to the higher extinction towards SN~2014J and the smaller telescope size of the NOT compared
to the LBT. For [Ca~II]~$\lambda\lambda$7291,7324 we estimate that at least $0.006~\msun$ of ablated material with solar-metallicity is 
needed to result in a detection. This is below the limit for SN~2011fe due to modest extinction for SN~2014J in the red, as well as 
a noisy region of the LBT spectrum at wavelenghts partly overlapping with those expected for the [Ca~II] lines.The upper mass limit on 
ablated masses from H$\alpha$, [O~I]~$\lambda6300$ and  [Ca~II]~$\lambda\lambda$7291,7324 correspond to observed line fluxes of 
$1.6\EE{-16}$~erg~s$^{-1}$~cm$^{-2}$, $1.3\EE{-16}$~erg~s$^{-1}$~cm$^{-2}$ and $3.8\EE{-16}$~erg~s$^{-1}$~cm$^{-2}$, respectively.

\begin{figure}
\setlength{\unitlength}{1mm}
\resizebox{19cm}{!}{
\begin{picture}(190,240)(0,0)
\put (0,175) {\includegraphics[width=91mm, angle=0,clip=]{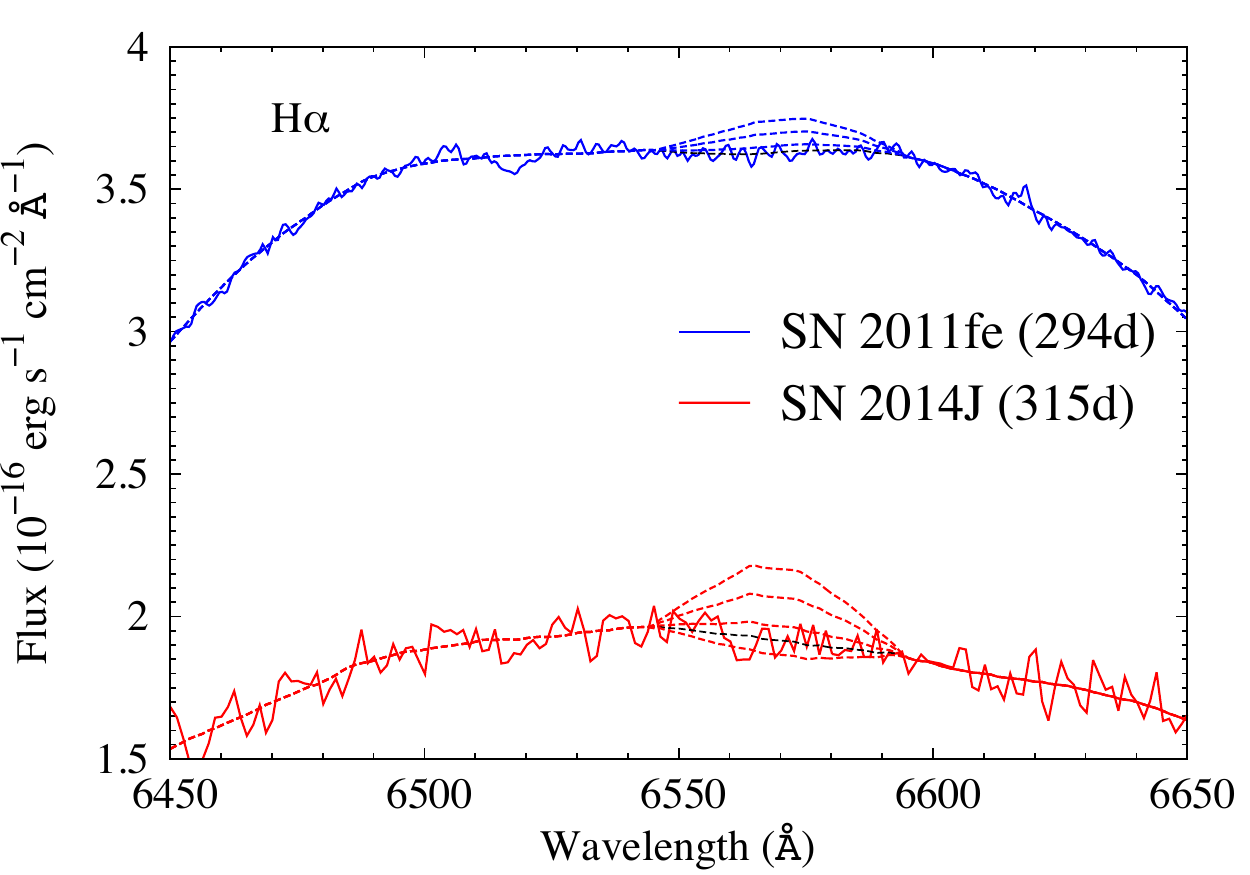}} 
\put (0,110) {\includegraphics[width=91mm, angle=0,clip=]{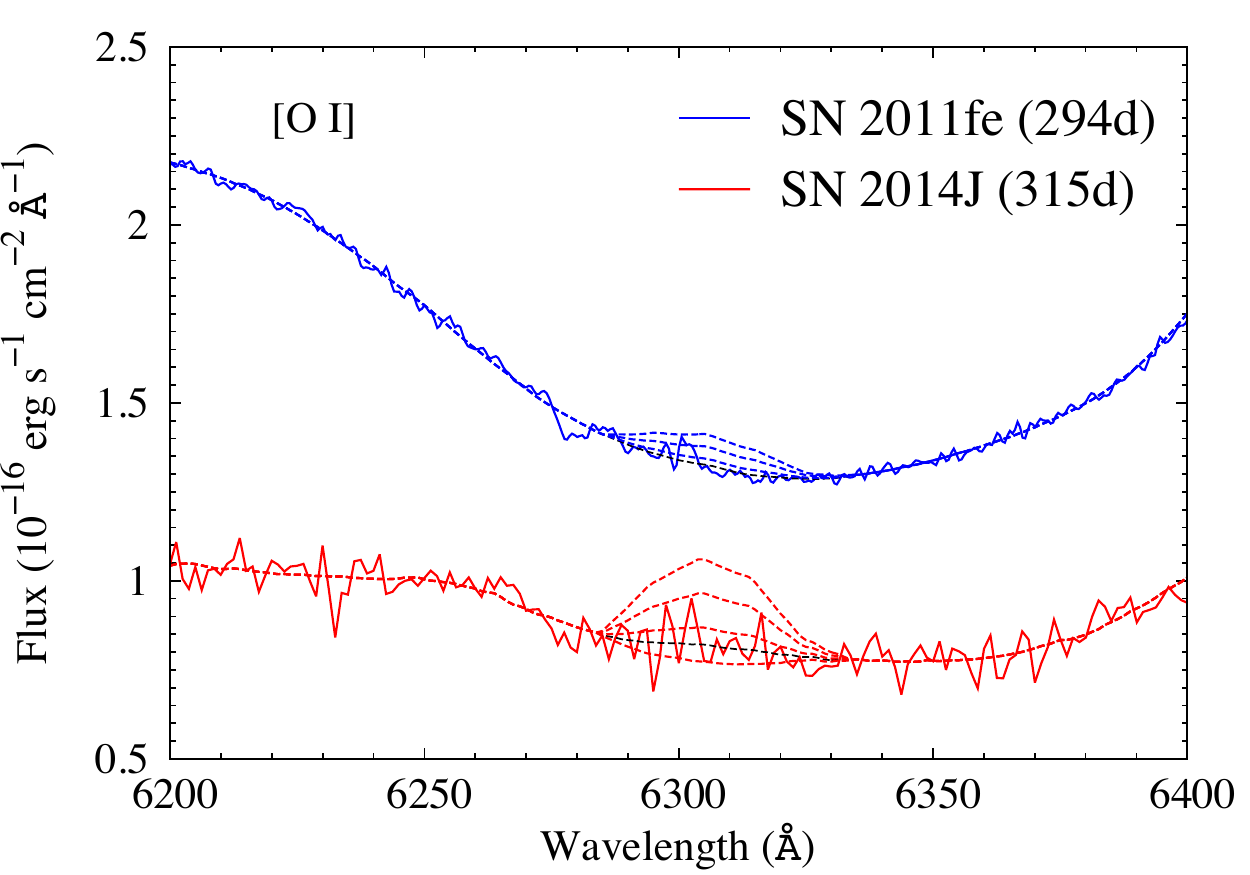}} 
\put (0,45) {\includegraphics[width=91mm, angle=0,clip=]{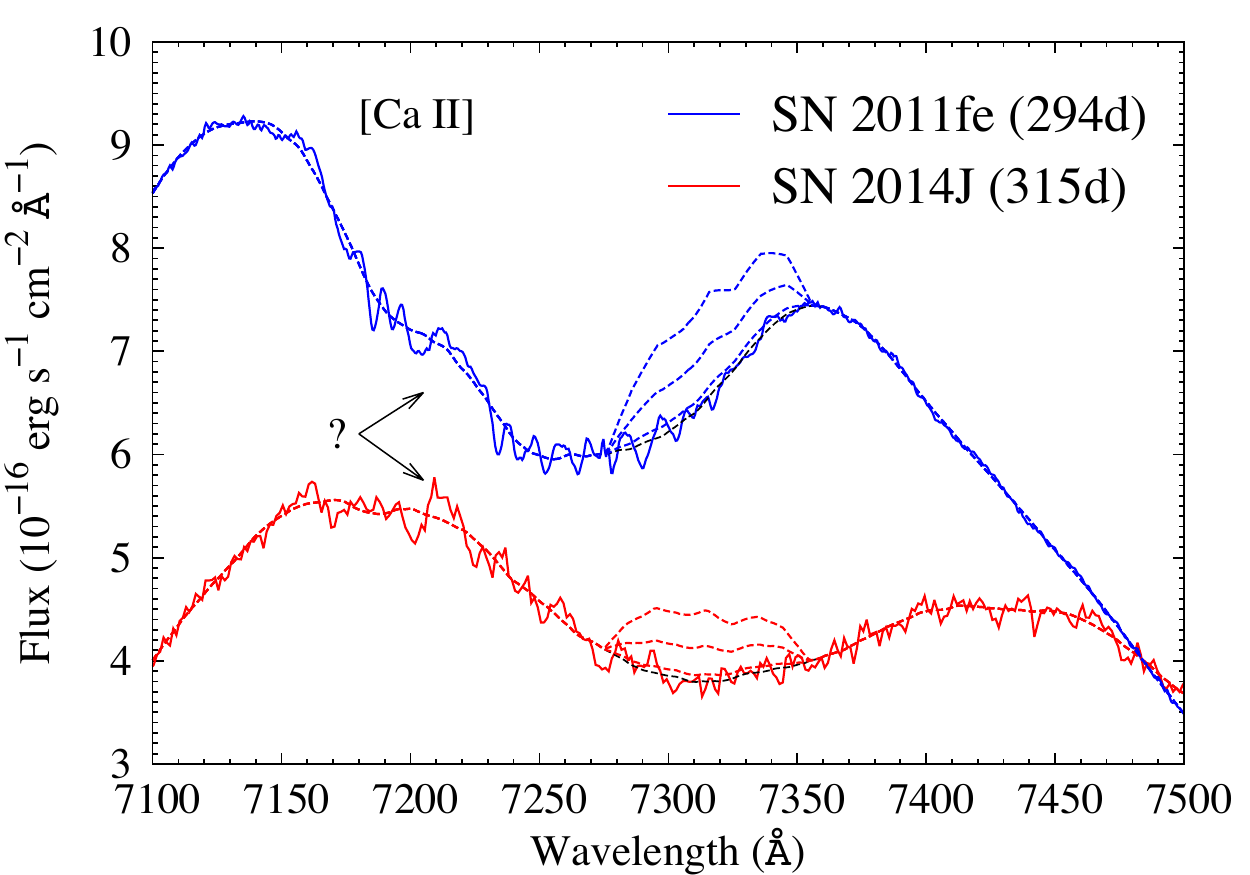}} 
\put (0,41) {\parbox[t]{88mm}{\caption{
Observed spectra (solid red and blue lines) with modelled line profiles (dashed red and blue lines) added. For H$\alpha$ (upper panel) 
and [O~I]~$\lambda6300$ (middle panel) the modelled fluxes are for 1$\times$, 3$\times$ and 5$\times$ the 3$\sigma$ statistical
limits on the masses estimated in Sect. 3.1 and 3.2. For H$\alpha$ and [O~I]~$\lambda6300$ we have also subtracted $1\times$ these 
masses (also shown by red dashed lines). For [Ca~II]~$\lambda\lambda$7291,7324 (lower panel) we have added modelled fluxes 
corresponding to
1$\times$, 5$\times$ and 10$\times$ the 3$\sigma$ statistical limits on the masses estimated in Sect. 3.1 and 3.2. The dashed lines 
(marked in black across the line profiles) tracing the full observed spectra are the smoothed spectra using Savitzky-Golay polynomials, 
as  outlined in Sect. 3.1  and 3.2. In the [Ca~II]~$\lambda\lambda$7291,7324 panel, we have highlighted a feature around 7210~\AA\ 
present in the spectra of both supernovae with a question mark. Note that the plots show observed wavelengths.
}\label{fig:sanity_check}}}
\end{picture}}
\end{figure}

\subsection{Sanity check of results}
The estimated 3$\sigma$ limits on ablated mass in Sect. 3.1 and 3.2 are strictly statistical. Additional systematic errors may arise 
due to our ignorance of the shape of the underlying spectrum from the supernova ejecta. The statistical approach artificially removes 
this uncertainty when the smoothed continuum is subtracted from the observed one. To highlight this, Fig. \ref{fig:sanity_check} shows 
the observed spectra, and compares that to smoothed spectra plus modelled line emission.

\subsubsection{H$\alpha$}
The upper panel of Fig. \ref{fig:sanity_check} shows the region around H$\alpha$ for both supernovae. The dashed lines that 
strike through the spectra are from the Savitzky-Golay-smoothed continuum approximation. At the position of the line, the
continuum is marked by a dashed black line. The dashed blue and red lines at that position are for 1$\times$, 3$\times$ and 5$\times$ 
the 3$\sigma$ statistical limits on the masses estimated in Sect. 3.1 and 3.2. For SN~2014J we have also added a dashed red
line for a subtraction of a spectrum corresponding to the 3$\sigma$ statistical limit. An inspection by eye shows that the 3$\sigma$ 
statistical limit could easily be taken as part of the supernova continuum for SN~2011fe. For a clear deviation from the general shape of 
the continuum, one should probably require 3$\times$ the 3$\sigma$ statistical limit for SN~2011fe, and somewhat less than 
3$\times$3$\sigma$ for SN~2014J, to also include systematic uncertainties. This means that the limit on ablated mass from  
H$\alpha$ should be $\sim 0.003~\msun$ for SN~2011fe and $\sim 0.0085~\msun$ for SN~2014J.
 
\subsubsection{[O~I]~$\lambda6300$}
The middle panel of Fig. \ref{fig:sanity_check} shows the region around  [O~I]~$\lambda$6300. The blue,
red and black lines have the same meaning as for the H$\alpha$ panel. An inspection by eye shows that 3$\times$ the 3$\sigma$ statistical limit is probably a safe upper limit for SN~2011fe, whereas 2$\times$ the 3$\sigma$ statistical limit should certainly be
enough for SN~2014J. This translates into upper limits on ablated mass using [O~I]~$\lambda$6300 to be $\sim 0.007~\msun$
for SN~2011fe and $\sim 0.02~\msun$ for SN~2014J.

\subsubsection{[Ca~II]~$\lambda\lambda7291,7324$}
The lower panel of Fig. \ref{fig:sanity_check} shows the region around  [Ca~II]~$\lambda\lambda$7291,7324 for both supernovae.
Here the dashed blue and red lines at that position are for 1$\times$, 5$\times$ and 10$\times$ the 3$\sigma$ statistical limits on 
the masses estimated in Sect. 3.1 and 3.2. For SN~2011fe even more than 5$\times$ the 3$\sigma$  statistical limit produces a spectrum 
which mistakenly could be part of the supernova continuum bump around $\sim 7350$~\AA, whereas 5$\times$ the 3$\sigma$ 
statistical limit probably is a fair upper limit on the ablated mass for SN~2014J. This corresponds to a limit on ablated mass from 
[Ca~II]~$\lambda\lambda$7291,7324 which is $\sim 0.06~\msun$ for SN~2011fe and $\sim 0.03~\msun$ for SN~2014J.

\section{Discussion}
The limits on ablated mass in Sect. 3 were derived under the assumption of solar abundance composition \citep{and89}. This is 
expected in a SD scenario with hydrogen-rich gas being stripped from the companion star. However, there is also the possibility of 
helium-dominated gas being stripped \citep{pan12,liu13b}. If the O/He and Ca/He ratios, and the efficiency of line emission in such 
a scenario, do not deviate significantly from the solar composition case, our results may provide rough upper limits on the ablated mass 
in case of a helium-rich donor. The upper limits from [O~I]~$\lambda$6300 and  [Ca~II]~$\lambda\lambda$7291,7324 will then 
be a factor of $4([X_{\rm He}/X_{\rm H}]/(1 + 4(X_{\rm He}/X_{\rm H})))$ lower than in Sect. 3. Here $X_{\rm He}/X_{\rm H}$ is the 
number density ratio of He and H for solar abundance. With $X_{\rm He}/X_{\rm H} = 0.085$, this factor becomes $\approx 0.25$, 
meaning that the upper limits on ablated mass from Sections 3.2.2. and 3.2.3. become $\sim 0.002~\msun$ and $\sim 0.005~\msun$
using [O~I]~$\lambda$6300 and $\sim 0.015~\msun$ and $\sim 0.008~\msun$ using [Ca~II]~$\lambda\lambda$7291,7324 for 
SNe~2011fe and 2014J, respectively. 

\begin{figure*}
\centering
\includegraphics[width=185mm,clip]{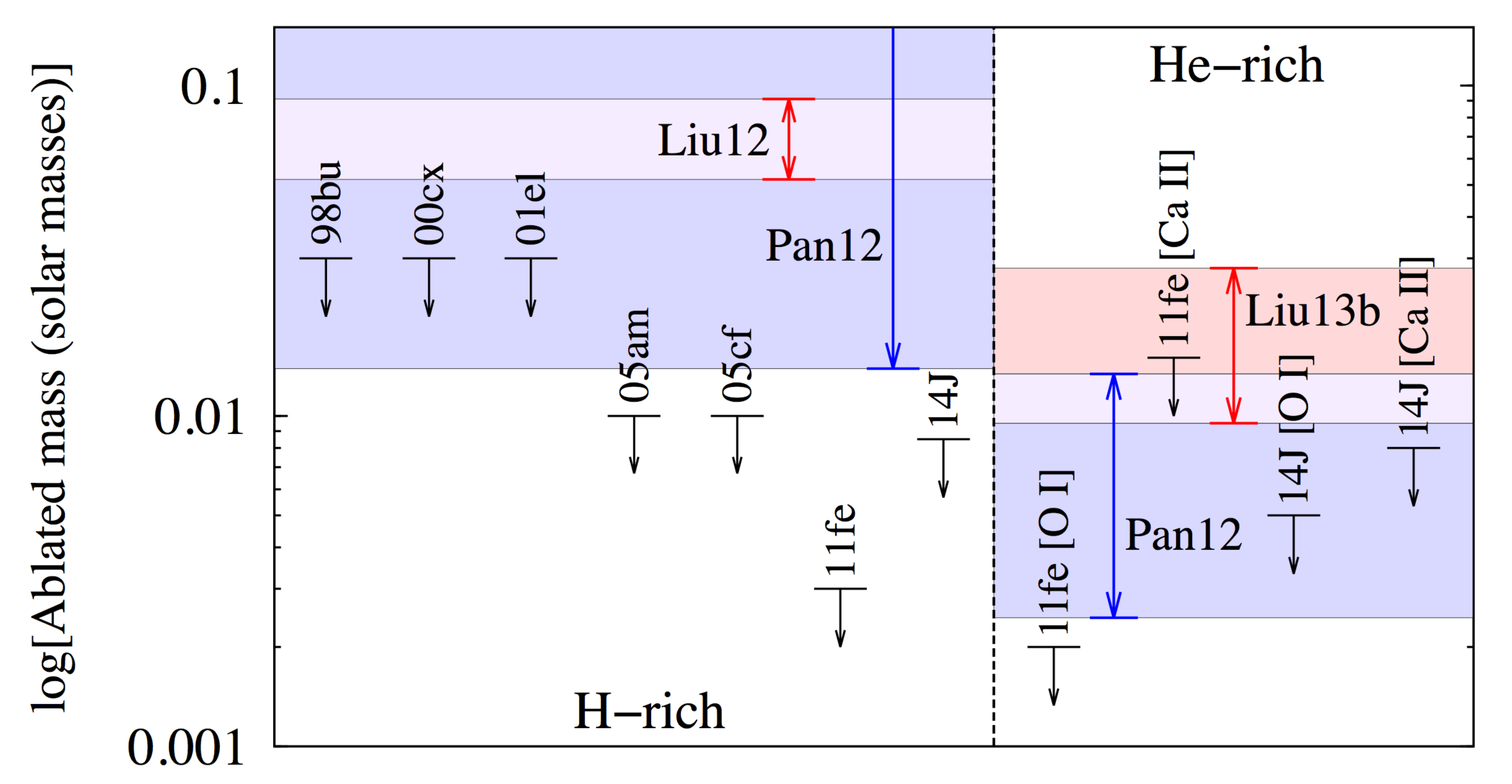}
\caption{Summary of estimated upper limits on the mass of ablated gas from a SD companion. The estimates for SNe~1998bu and
2000cx are from \citet{lun13}, the one for SN~2001el from \citet{mat05}, the ones for SNe~2005am and 2005cf from \citet{leo07}, 
and the ones for SNe~2011fe and 2014J are from this paper. The left part of the figure is for hydrogen-rich gas, and the right part
is for helium-dominated gas. Simulated ranges of ablated mass are marked by filled areas and arrows. For hydrogen-rich donors, red 
is for \citet{liu12} and blue is for \citet{pan12}, while for helium-dominated donors, red is for \citet{liu13b} and blue is for \citet{pan12}. 
Note that our limits for both SNe~2011fe and 2014J are below the simulated ranges for hydrogen-rich gas, while helium-rich donors
cannot be fully ruled out, in particular not for SN~2014J, if we are guided by the simulations of \citet{pan12}. See text for further details. 
              }
\label{fig:ablate}
\end{figure*}

In Fig. \ref{fig:ablate} we summarise the predicted ablated masses and the upper limits from observations. We include both hydrogen- 
and helium-rich donors, where the predicted ablated masses for hydrogen-rich donors were taken from \citet{pan12} and those expected 
in the helium-rich case from \citet{pan12} and \citet{liu13b}. For the results of \citet{pan12}, we have used the information in their Table~2, 
while for the results of \citet{liu12,liu13b} we follow the advice in \citet{liu13b} to assume that $50\%$ of the gas lost from the companion 
due to supernova impact is ablated. The mass range of ablated hydrogen-rich gas is therefore $0.052-0.091~\msun$ according to
\citet[][which supersedes the models of Pakmor et al. 2008]{liu12} or $0.0139-0.636~\msun$ according to \citet{pan12}, while for 
the helium-rich case, we have in Fig. \ref{fig:ablate} marked the ranges $0.00245-0.0134~\msun$ \citep{pan12} 
and $0.0095-0.028~\msun$ \citep{liu13b}.

Our limits are sensitive to the velocity of the stripped/ablated gas. We have assumed that the ablated gas is confined
to $10^3 \kms$. This assumption agrees with the results of \citet{pan12}, where the peak of the velocity distribution is well below
$10^3 \kms$ for hydrogen-rich donors, but just short of this velocity for helium-rich donors \citep[see also][for a confirmation]{liu13b};  
there is a significant fraction of gas with velocities in the range $(1-2)\times10^3 \kms$ in the helium-donor case. A higher velocity
than our assumed $10^3 \kms$ results in shallower emission line profiles from the ablated gas, and the limits on 
ablated gas for helium-rich donors should therefore be shifted upwards in Fig. \ref{fig:ablate}. Judging from the velocity distribution
of the ablated gas in \citet{pan12}, the upward shift could be a factor of $\sim 1.3-1.5$, whereas for hydrogen-rich donors the factor 
could instead be shifted in the other direction by a similar amount since the expected velocity of the ablated gas is well below 
$10^3 \kms$ for such donors.

\subsection{Implications for SNe~2011fe and 2014J}
\subsubsection{Constraints from our findings}
From Fig. \ref{fig:ablate} it is obvious that hydrogen-rich donor stars are disfavoured for both SNe~2011fe and 2014J. \citet{pan12}
made models for both red giant and main sequence companions, and the main sequence stars populate the lower part of the 
range marked `Liu12'  in Fig. \ref{fig:ablate}. Red giant hydrogen-rich companions for SNe~2011fe and 2014J are therefore clearly ruled
out in terms of ablated  mass. The models with the lowest amount of ablated gas are those with main-sequence companions, and among
these the ablated mass decreases with increasing orbital separation. The largest separation tested by \citet{pan12} was 
$2.75\EE{11}$~cm, corresponding to $5R_\star$, where  $R_\star$ is the donor star radius. They found a power-law relation between the 
separation (in units of $R_\star$) and the amount of unbound matter from the donor. Extrapolating their grid of models, the 
separation would need to be $\gsim 6~R_\star$ for the ablated mass to be as low as our upper limit for SN~2014J, and 
$\gsim 8.5~R_\star$ for SN~2011fe. Models with such large separation were tested by \citet{pak08},who found that the stripped
mass could be below $0.01~\msun$ for models with large separation, and even lower if the explosion energy is reduced. Although the 
trends are clear, the exact numbers in \citet{pak08} are uncertain, as cautioned by \citet{liu12} and \citet{pan12}.

While the models explored by \citet{pan12} used binary evolution models of \citet{iva04} as input, \citet{liu12} modelled both the binary 
evolution and simulated the explosion impact themselves. \citet{liu12} concentrated on main sequence companions, and, like
 \citet{pan12},
they found a clear trend of  decreasing ablated mass for increasing binary separation when expressed in $R_\star$. The ablated mass 
is, however, larger than in the models of \citet{pan12}, and our mass limits on ablated hydrogen-rich gas for both SNe~2011fe and 2014J
are much lower than in the models of \citet{liu12}. It remains to be tested which of the impact simulations are the most accurate in terms 
of ablated mass. In any case, a large separation is necessary to make the impact models compatible with our
limits on ablated hydrogen-rich gas.

If the SD companion were instead a helium-rich donor, the right part of Fig. \ref{fig:ablate} shows that it is only our limits for SN~2011fe 
which are below the expected mass of ablated gas from impact models. However, if one adjusts for a likely slightly larger velocity of the
ablated gas for helium-rich donors than assumed in the construction of Fig. \ref{fig:ablate}, we cannot rule out all models of helium-rich 
donors in  \citet{pan12} even for SN~2011fe. For SN~2014J, our upper limits from both  [O~I]~$\lambda6300$ and 
[Ca~II]~$\lambda\lambda$7291,7324 are higher than the lower part of the range of ablated masses in the models of \citet{pan12}. Like for 
the hydrogen-rich case, it is the systems with the largest separation that produce the least amount of helium-rich ablated gas in these 
models. The helium-rich companion in \citet{pan12} is from \citet{wan09}, and not from simulations of binary evolution. There could 
therefore be some uncertainty regarding the binary evolution, and thus perhaps the amount of ablated gas. \citet{liu12}, on the other 
hand, follow the detailed binary evolution leading up to the explosion. The smallest amount of unbound material from the companion 
occurs for the systems with the largest separation. Like for hydrogen-rich companions, there is an inconsistency between the 
impact models of Liu and co-workers and \citet{pan12} with regard to the amount of ablated mass. Until this is settled, it cannot be 
ruled out that the progenitor system of SN~2014J, and perhaps even that of SN~2011fe, could have been a WD with a helium-rich 
non-degenerate companion at large separation. 

\subsubsection{Checking against other constraints on SNe~2011fe and 2014J}
Pre-explosion imaging of SN~2011fe \citep{li11} cannot fully rule out a helium-rich donor as a possibility for the origin of the system. 
Donors with $M_V \gsim -0.6$~mag, in combination with $T_{\rm eff} \gsim 50\,000$~K are allowed. Here $M_V$ is the 
absolute visual magnitude of the donor at the time of explosion. Using the bolometric corrections of \citet{tor10}, the bolometric 
luminosity of a tentative helium-rich donor in the SN~2011fe progenitor system was log$(L_{\rm bol}/L_{\odot,{\rm bol}}) \lsim 3.5$. 
Most of the helium-donor stars in \citet{liu13b}, and \citet[][who made a thorough investigation of helium-donor systems]{wan09} have 
properties in this range, so our analysis (cf. Fig. \ref{fig:ablate}) could be more constraining than the pre-explosion imaging, especially if 
the impact models of \citet{liu13b} are closer to the real situation than those of \citet{pan12} in terms of ablated gas. 

Related to our results for SN~2011fe is the {\it Swift} study by \citet{bro12}. The non-detection of very early ultraviolet emission from 
the supernova limits the parameter space of allowed SD companions to only include main-sequence companions with masses 
$\lsim 2-3.5~\msun$, and perhaps even $\lsim 1~\msun$, for close companions. Geometric probabilities are less than 1\%
for a $2~(1)~\msun$ main-sequence star separated from the white dwarf by $5 ( 3) \EE{11}$~cm, i.e., close to the largest separation
of $3\EE{11}$~cm tested by \citet{pan12} for hydrogen-rich companions, and decreases further for larger separation. Our analysis 
for SN~2011fe, in combination with the models of \citet{pan12} rules out a SD system with a hydrogen-rich donor at a separation of 
$\lsim 8.5~R_\star \sim 4\EE{11}$~cm. Adding the constraints from the analysis of the {\it Swift} observations essentially rules out all 
SD scenarios with a main-sequence hydrogen-rich donor for SN~2011fe. A $2-3~\msun$ main-sequence could be possible if it lies
within a few degrees along the line-of-sight on the rear side of the white dwarf. Such a star would also pass the limits set by the
pre-explosion imaging  \citep{li11}, but is likely at odds with the very early optical observations discussed by \citet{bloo12}. 

The only hydrogen-rich SD scenario reasonably possible for SN~2011fe is that of a spun up/spun down super-Chandrasekhar WD 
\citep{dis11,jus11,hachisu12}. In such systems the donor star shrinks far inside its Roche lobe prior to the explosion, making the SD 
companion smaller and more tightly bound. The supernova ejecta impact on such a star should also produce low enough emission to 
pass the limits from the early UV observations of \citet{bro12}. Furthermore, very small amounts of ablated gas are expected, as well 
as very dilute ($n \sim 1 \cm3$) circumstellar gas in the vicinity of the supernova. A way to constrain this scenario is through continued 
deep monitoring of the supernova in radio \citep{per14}.

Our results for SN~2011fe are also consistent with previous findings \citep[cf.][]{mao14} that it may indeed have been the outcome of 
a DD scenario. What speaks in favour of a DD scenario rather than the spun up/spun down super-Chandrasekhar WD scenario is that 
SN~2011fe was a normal SN~Ia in terms of lightcurve and spectral evolution, and that the spun up/spun down super-Chandrasekhar WD 
scenario is not thought to be the normal path leading to a SN~Ia explosion. However, while the DD scenario for SN~2011fe may be likely, 
we cannot rule out a spun up/spun down super-Chandrasekhar WD. Neither can we fully rule out a helium-rich donor at large separation 
(cf. above). Using the numbers in Fig. \ref{fig:ablate}  and figure 12 in \citet{pan12}, we estimate that we cannot rule separations
which are $\gsim 6~R_\star \sim 8\EE{10}$~cm and $\gsim 4.5~R_\star \sim 6\EE{10}$~cm for helium-rich donors for SNe~2011fe
and 2014J, respectively. Even if not explicitly discussed by \citet{bro12}, helium-rich donor systems with such small separation may not be
well constrained by the early $\it Swift$ observations of SN~2011fe.

Recently, \citet{tau14} reported very late (1034 days past explosion) observations of SN~2011fe. No sign of narrow H$\alpha$
was found, but this could be due to poor signal-to-noise. A tentative identification of [O~I] was, however, made, which would make
SN~2011fe the third SN~Ia ever, besides SNe~1937C \citep{min39} and the subluminous 2010lp \citep{tau13}, to show signs of 
[O~I] in late spectra. The [O~I] emission in SN~2010lp is unlike that we expect from ablated gas since the line profiles indicate
emission at velocities offset by $\sim 1900 \kms$ from the rest wavelengths of [O~I]~$\lambda\lambda$6300,6364. If the [O~I] identification is correct for SN~2011fe, then a similarly large offset ($\sim+2000 \kms$) would apply. The [O~I] emission in these SNe 
has therefore probably nothing to do with ablated gas from a companion, but should come from blobs of oxygen-rich SN ejecta. In the
case of SN~2011fe, we emphasise that the [O~I] identification could also be a misinterpretation, as the emission may very well be due 
iron \citep{tau14}.

Just as for SN~2011fe, the progenitor system for SN~2014J, could have been a DD system, a spun-up/spun-down 
super-Chandrasekhar WD scenario, or a system with a helium-rich companion at large separation. For SN~2014J, our results could
also be compatible with a well-separated hydrogen-rich donor system. A recent clue, perhaps in favour of a SD scenario, was the 
reported early variation in two narrow absorption components of K~I~$\lambda$7665 \citep{gra14}. The estimated distance from the 
supernova to the absorbing gas is, however, $\sim 10^{19}$~cm, which is $\sim 50-100$ times further away from the supernova than 
the likely position of a circumstellar blast wave after 1 year \citep{per14}. Continued monitoring in radio may pick up circumstellar gas 
closer to the supernova. Unlike narrow absorption lines, radio is sensitive to any gas close to the supernova, and not only gas along 
the line of sight to it. 

Further clues to the origin of SN~2014J come from very early photometry of the supernova. As reported by \citet{goo14b}, the
rise in luminosity during the first few days after explosion indicates an extra energy source which could be due to interaction of the 
ejecta with a non-degenerate companion \citep[cf.][]{kas10} or the debris from a disrupted WD \citep[e.g.,][]{lev14}. The matter interacting 
with the ejecta must be confined to the immediate vicinity of the explosion site since radio observations only $8-9$ days after the
explosion did not reveal any emission \citep[][see also P\'erez-Torres et al. 2014]{Chandler2014}.

Analysis of pre-explosion
images \citep{kel14} shows that a DD progenitor system, a helium-star donor with low effective temperature ($T_{\rm eff}$), 
or a system like U Sco (recurrent nova and a supersoft-X-ray source with a subgiant companion) would not show up in the 
pre-explosion images. On the other hand, SD systems like V445 Pup (bright helium-star donor) or RS Oph (bright recurrent nova 
and a symbiotic source) are both excluded. 

More specifically, the pre-explosion imaging of SN~2014J cannot rule out helium-star donors with $M_V \gsim -2.5$~mag in 
combination with $T_{\rm eff} \gsim 40\,000$~K \citep[cf. Figure 4 of][]{kel14}. Again, using the bolometric corrections of \citet{tor10}, 
the bolometric luminosity of a tentative helium-rich donor in the SN~2014J progenitor system was 
log$(L_{\rm bol}/L_{\odot,{\rm bol}}) \lsim 4.4$. The helium-donor stars in \citet{liu13b} and \citet{wan09} all have properties in 
this range, so our analysis (cf. Fig. \ref{fig:ablate}) is more constraining than the pre-imaging for helium-donor stars, in particular if 
the simulations of \citet{liu13b} are more representative than those of \citet{pan12}. We note, however, that \citet{liu13b}  assume 
that the metal abundance of the helium-rich donor remains at $Z = 0.02$ (i.e., the solar value) even when hydrogen has been removed, 
whereas we have in Fig. \ref{fig:ablate} assumed that this number is a factor of 4 higher for helium-rich donor compared to hydrogen-rich 
ones. If we abandon this correction factor, and also consider the slightly larger velocity of the ablated gas in the helium-rich scenario 
than assumed in Fig. \ref{fig:ablate}, the ablated mass in the models of \citet{liu13b} become consistent with our results, in particular for 
the helium-star donors with the largest separation to the WD. Removing the correction factor of 4 could make it easier
to accommodate a helium-star donor system also for SN~2011fe. For SN~2014J, we note that a helium-rich donor was argued for by 
\citet{diehl14} to explain the early emergence of gamma-ray line emission.

\subsection{Broad lines of SN~2011fe and 2014J}
The main spectral peak between 7050--7250 \AA\ for SN~2014J is centered at $\approx 7170$~\AA, whereas it is shifted to the 
blue at $\approx 7135$~\AA\ for SN~2011fe. This is close to the rest-wavelength of [Fe~II]~$\lambda$7155. Likewise, 
the main peak between 7250--7300 \AA\ and 7500 \AA\ is clearly shifted more to the blue for SN~2011fe, where the peak 
occurs at $\sim 7355$~\AA, compared to $\sim 7420$~\AA\ for SN~2014J. Most of this peak can be attributed to 
[Ni~II]~$\lambda$7378. Correcting for the redshifts of the SNe, and assuming that [Fe~II]~$\lambda$7155 and [Ni~II]~$\lambda$7378 
are the main contributors \citep[see also][]{mae10b,tau14,gra15}, the [Fe~II]~$\lambda$7155 peak occurs at $\sim -1000 \kms$ and
$\sim +400 \kms$ for SNe~2011fe and 2014J, respectively, whereas for [Ni~II]~$\lambda$7378 they are at $\sim -1100 \kms$ and
$\sim +1300 \kms$ for SNe~2011fe and 2014J, respectively. There is thus a consistent blueshift for SN~2011fe, which is in full 
agreement with \citet[][see also Graham et al. 2015]{mcc13}, whereas the lines are redshifted for SN~2014J. 
Figure \ref{fig:14J_spectrum} does not show any obvious similar shifts between the supernovae for the broad peaks with centers 
around 5900~\AA\ and 6550~\AA, which are thought to be dominated by [Co~III] lines \citep{mae10b,tau14,gra15}. This is in agreement 
with the analysis of \citet{mae10b} where [Fe~II]~$\lambda$7155 and [Ni~II]~$\lambda$7378 mainly originate from the dense central 
parts of the ejecta where asymmetries can be expected, as opposed to the more highly ionized exterior region, which is closer to 
being spherically symmetric. \citet{mae10b} argue that this is a natural outcome of a delayed-detonation scenario, and exemplify this 
for 12 SNe~Ia. Half of them have shifts of [Ni~II]~$\lambda$7378 in excess of $1500 \kms$. Both SNe~2011fe and 2014J have 
smaller velocity offsets, which may just be a viewing angle effect.

\citet{mae10a} further connect the viewing angle to how fast the absorption trough of Si~II~$\lambda6355$ recedes after $B-$band
maximum. The decline rate in this velocity ($v_{\rm SiII}$) is called $\dot v_{\rm Si}$, and \citet{mae10a} argue that SNe~Ia 
with $\dot v_{\rm Si} \gsim 70 \kms$~d$^{-1}$ \citep[the so-called high-velocity gradient, or HVG, group, cf.][]{ben05} all have redshifted 
[Fe~II]~$\lambda$7155 and [Ni~II]~$\lambda$7378 emission in nebular spectra. There is also a small fraction of those SNe~Ia
with $\dot v_{\rm Si} \lsim 70 \kms$~d$^{-1}$ (the so-called low-velocity gradient, or LVG, group) that have redshifted nebular
lines, but the majority of the LVG group SNe~Ia have blueshifted [Fe~II]~$\lambda$7155 and [Ni~II]~$\lambda$7378 emission.

Consulting the results of \citet{kaw14} and \citet{mari15} for SN~2014J, we find that $\dot v_{\rm Si} \sim 55 \kms$~d$^{-1}$ 
(between $0-30$ days after $B$ maximum) and  $\dot v_{\rm Si} \sim 50 \kms$~d$^{-1}$ (between $-0.7$ and $+9.3$ days after $B$ maximum), respectively. Furthermore, \citet{ash14} found $\dot v_{\rm Si} \approx 58.8 \kms$~d$^{-1}$ between $0-10$ days 
after $B$ maximum using their observations. This would put SN~2014J in the LVG group, and the supernova would belong to the 
minority of SNe~Ia which are both LVG group objects and have redshifted nebular [Fe~II]~$\lambda$7155 and 
[Ni~II]~$\lambda$7378 emission. Another one in this category is SN~2001el \citep[][this SN is further discussed in Sect. 4.4]{mae10a}.
We note that no LVG SNe in \citet{ben05} have 
$v_{\rm SiII} \geq 11000 \kms$ around $B-$band maximum, whereas $v_{\rm SiII} \approx 11750 \kms$ for SN~2014J at that epoch 
\citep{ash14,kaw14,mari15}. 

For the LVG group SN~2011fe \citep{par12,gra15}, the blueshifted nebular [Fe~II]~$\lambda$7155 and 
[Ni~II]~$\lambda$7378 emission fits well into the model of \citet[][see also McClelland et al. 2013]{mae10a}. SN~2011fe had also 
notably lower $v_{\rm SiII}$ than SN~2014J around $B-$band maximum \citep{par12,goo14a}.

\subsection{The 7210~\AA\ feature in the spectra of SN~2011fe and 2014J}
As indicated in Fig. \ref{fig:sanity_check}, there appears to be a spectral feature around the observed wavelength 7210~\AA\ for both 
SN~2011fe and 2014J, with a width roughly like that expected from a single spectral line from ablated gas. Had this feature coincided 
with the expected wavelength of, e.g., H$\alpha$, the flux of the feature would have corresponded to a level greater than 
the $3 \sigma$ statistical limit of the H$\alpha$ flux in Fig. \ref{fig:14J_spectra}, and could have mistakenly been taken as
evidence of ablated gas. This shows the importance of making a sanity check like that in Sect. 3.3, and not just relying on statistical 
errors to set upper limits on spectral line fluxes. The question arises whether the feature is due to clumpiness/asymmetry in the 
supernova ejecta, ablated gas from a companion, some other source, or if it is an observational artefact. The feature
is seen in all four individual frames for SN~2014J and appears in both SNe. From an inspection of Fig. \ref{fig:14J_spectrum}, 
the feature appears to be the only one of its sort, except perhaps for a feature at 7155~\AA\ in the SN~2014J spectrum.
The spectra of SN~2011fe at 329 days by \citet{gra15}  and at 331 days by \citet{tau14} have, unfortunately, too low 
signal-to-noise to support or reject the tentative feature at 7210~\AA\ for that SN.

We have used The Atomic Line List V2.05B18\footnote{http://www.pa.uky.edu/$\sim$peter/newpage/} to search for possible spectral lines 
that could explain the 7210~\AA\ feature of both SNe in Fig. \ref{fig:sanity_check}, but find no obvious counterpart other than those
most likely responsible for the main peaks, i.e., the usual suspects of forbidden lines of Fe and Ni \citep[e.g.,][]{mcc13}. Neither do we 
find any other obvious candidate for the 7155~\AA\ feature. We have also looked at late spectra of SNe 1998bu, 2000cx, 
2001el and 2005cf (cf. Fig. \ref{fig:ablate} for references discussing these spectra), but do not find a similar feature for those SNe. 
This could partly be due to lower signal-to-noise in the spectra of these SNe. In any case, there is no support for a 7210~\AA\ feature in 
their spectra. 

The 7155~\AA\ and 7210~\AA\ features occur in a spectral region with telluric molecular absorption, which can also be traced in the 
standard stars used during the LBT and NOT runs. The most likely explanation for the features is that they are therefore 
observational artefacts from this absorption.

\subsection{Implications for SNe 1998bu, 2000cx, 2001el, 2005am and 2005cf}
Apart from SNe~2011fe and 2014J, Fig. \ref{fig:ablate} shows upper limits on hydrogen-rich ablated gas for the five previous SNe~Ia 
for which there are upper limits on H$\alpha$ in late spectra. These limits are all lower than in the models of \citet{liu12}, and are also 
close to the lowest mass of  hydrogen-rich ablated gas in the models of \citet{pan12}. No estimated limits on helium-rich ablated gas 
exist for these SNe. Possible progenitor models for these five SNe could therefore be helium-rich donor systems, DD
systems, spun-up/spun-down super-Chandrasekhar WD progenitors, or perhaps hydrogen-rich donor systems with a large
separation between the WD and the non-degenerate companion, likely being a main-sequence star as donor \citep[cf. the models of][]
{pan12}. The only systems fully ruled out are those with red giant donors, and those with close main-sequence donors. According
to figure 12 of \citet{pan12}, main-sequence donor systems with a separation of $\lsim 6 R_\star$ are ruled out for SNe~2005am and
2005cf, as well as $\lsim 4.5 R_\star$ for SNe 1998bu, 2000cx, 2001el. 

There are limited constraints on the progenitor systems of these supernovae from other investigations. For example, narrow emission
lines were looked for in spectra of SNe~2000cx and 2001el, but no such emission was detected \citep{mat05,lun13}, and 
SN~2000cx showed no time-varying narrow interstellar/circumstellar absorption features \citep{pat07a}. SN~1998bu had a light 
echo \citep{cap01,gar01}, but that was due to foreground material, and not to circumstellar matter. SN~2005cf was observed early 
in the ultraviolet (UV) with the $\it HST$ \citep{wan12} and with $\it Swift$, starting $-8.8$ and $-7.8$ days 
before $B$ maximum, respectively. A comprehensive compilation of data and an analysis were presented in \citet{gal12}, but there is no 
evidence of enhanced early flux indicative of the ejecta interacting with nearby material.  $\it Swift$ also observed SN~2005am in the 
UV from $-1$ day, and onwards \citep{buf09}, but nothing conspicuous with regard to the nature of the progenitor system was detected.

Among all these SNe, SN~2000cx is clearly the oddball. Although it shares some properties with the overluminous SN~1991T, it
is different enough to form a separate category together with SN~2013bh \citep{sil13}. It probably stems from an old, low-metallicity 
population, which together with its spectral evolution indicate a DD or a SD delayed detonation scenario 
\citep[][and references therein]{sil13}. As a helium-star donor system is likely to originate from relatively massive progenitors, 
\citet{wan09} estimate that the maximum delay time for these SNe is $\sim 10^8$ years. This probably rules out such a progenitor
for SN~2000cx.

\subsection{Uncertainties}
In our models we have used the W7 model \citep{nom84} which produces $0.6~\msun$ of $^{56}$Ni. The excitation of the ablated
gas depends on the exact amount of $^{56}$Ni, as well as the distributions of the nickel and the ablated gas. It also depends on where
the positrons deposit their energy. As discussed in \citet{sol04}, we have assumed local and instantaneous deposition of the positron
energy. Neither microscopic nor macroscopic mixing of the ablated gas into the supernova ejecta was made. How all this is included 
and treated in our models affects the predicted fluxes of the lines we discuss. However, none of these 
uncertainties should translate into dramatic changes with regard to the modelled line emission. 

Of potentially greater importance is the fact that the number of elements and atomic levels in our models are somewhat 
limited  \citep[cf.][and references therein]{sol04}. This could make us 
underestimate line scattering and fluorescence. As noted in \citet{per14}, more recent models \citep[e.g.,][]{jer11} with more extensive 
line lists and more complete sets of ions and atomic levels, albeit not yet time-dependent, should be used to estimate these effects.

As we have remarked, the amount of ablated gas, or rather, the gas lost by a SD companion at velocities $\lsim 10^3 \kms$ after 
impact differs between models from different research groups. This is highlighted by Fig. \ref{fig:ablate}. Further such modelling is 
warranted. In 
particular for the fairly restricted range of possible progenitor systems of SNe~2011fe and 2014J (cf. Sect. 4.1). As an important boundary 
condition for possible SD progenitor systems, one must consider the strengthened evidence of a fairly long ($10^8$ yrs) minimum delay 
time for SNe~Ia in general \citep[e.g.,][]{and14}. This could prove hazardous for the helium-star donor channel since the maximum delay
time for such systems could be $\sim 10^8$ yrs \citep{wan09}. Refined binary evolution models are needed to see whether this channel
is likely to produce a noticeable fraction of SNe~Ia.

There is also uncertainty in our results due to the adopted distance and reddening to the supernovae. According to 
NED\footnote{http://ned.ipac.caltech.edu}, the uncertainty in distance to SN~2011fe is $\sim 6$\% and for modern measurements to 
M82 (for SN~2014J) it is $\sim 9$\%. If we assign 10\% as an uncertainty for the distance in general, this transforms into $\sim 20$\% in 
estimated values for the ablated mass. For SN~2011fe, uncertainties due to reddening is not an issue, whereas for SN~2014J the 
reddening is significant. There is good knowledge about the absolute luminosity so the reddening is well established \citep{ama14}. 
The combined effect of distance and reddening is estimated to cause an uncertainty in the ablated mass of $\sim 30$\% for SN~2014J. 
This is not insignificant, but smaller than the uncertainties in our modelling. 

\section{Conclusions and outlook}

We have observed SN~2014J with NOT/ALFOSC at 315 days after the explosion, and also used an archival spectrum of SN 2011fe 
294 days past explosion \citep[presented in][]{sha13b} to see if there is any trace of ablated gas from a SD companion. Guided by our 
modelling in \citet{lun13}, we have concentrated on possible emission in H$\alpha$, [O~I]~$\lambda$6300 or 
[Ca~II]~$\lambda\lambda$7291,7324.
We find no such emission, and from that we derive statistical upper limits on the mass of hydrogen-rich gas. These limits are, however, 
shown to be overwhelmed by systematic effects. When the latter are included, the limits on hydrogen-rich ablated gas are 
$0.003~\msun$ and $0.0085~\msun$ for SNe~2011fe and 2014J, respectively, where the limit for SN~2011fe should supersede that of
\citet{sha13b}, and that for SN~2014J is the second lowest ever. Assuming that the O/He and Ca/He ratios, and the efficiency of line
emission, are the same for helium-dominated ablated gas and for hydrogen-rich gas, we derive upper limits on the mass of 
helium-rich ablated gas. In this case, [O~I]~$\lambda$6300 provides the most stringent upper limits on the ablated gas, which are
$0.002~\msun$ and $0.005~\msun$ for SNe~2011fe and 2014J, respectively. 

These upper limits are compared with the most recent models predicting the amount of stripped and ablated gas presented by \citet{liu12,liu13b} and \citet{pan12}. For hydrogen-rich donors, our results are incompatible with red giants, and with main-sequence
donors if the separation between the binary companions are $\lsim 6~R_\star$ for SN~2014J and $\lsim 8.5~R_\star$ for SN~2011fe,
where $R_\star = 5.51\EE{10}$~cm is the radius of the main-sequence companion in the models of \citet{pan12}. Also, most helium-rich
donors are ruled out, except for those with the largest separation. Using the models of \citet{pan12}, helium-rich donors with a
separation of $\lsim 8\EE{10}$~cm and $\lsim 6\EE{10}$~cm from the white dwarf are ruled out for SNe~2011fe and 2014J,
respectively.

When we combine these results with findings from pre-explosion imaging and very early observations constraining possible interactions
with a companion, accretion disk or circumstellar matter, then essentially all hydrogen-rich main-sequence
donor systems can be ruled out for SN~2011fe, while we cannot state this for SN~2014J.  For both supernovae, our results are the
most constraining so far for helium-rich donors. Helium-rich donor systems may, however, have other problems, since the likely
modelled delay time of such systems is $\lsim 10^8$ years \citep{wan09}, at the same time as recent observational findings of 
SNe~Ia show a general delay time of $\gsim 10^8$ years \citep{and14}, leaving only a low probability of these systems to be progenitors
of SNe~Ia in general. Some support for a single-degenerate origin of SN~2014J could come from very early observations showing 
enhanced emission compared to the expected one \citep{goo14b} for an isolated exploding white dwarf, as well as the early emergence
of gamma-ray line emission \citep{diehl14}. According to our findings, the 
tentative non-degenerate companion would have to be well separated from the white dwarf. Other possible 
progenitor systems for SNe 2011fe and 2014J are single-degenerate systems with a spun up/spun down super-Chandrasekhar white 
dwarf, or double-degenerate systems. Continued radio monitoring of the SNe may reveal if any of these systems could be possible 
\citep{per14}. 

Data for SNe~1998bu, 2000cx, 2001el, 2005am and 2005cf are used to constrain their origin. Possible 
progenitor models for these SNe are found to be helium-rich donor systems, DD systems, spun-up/spun-down super-Chandrasekhar 
WD progenitors, or perhaps systems with a main-sequence star as donor, provided they are well separated from the white dwarfs. 
However, as for SNe~2011fe and 2014J, helium-rich donors could be rather unlikely due to their short delay times, and is probably 
excluded for SN~2000cx due to the nature of its host galaxy. 
  
For the broad lines of SNe~2011fe and 2014J it is found that the [Ni~II]~$\lambda$7378 emission is redshifted by $\sim +1300~\kms$, 
as opposed to a blueshift of $\sim -1100~\kms$ for SN~2011fe. Also, [Fe~II]~$\lambda$7155 appears to be redshifted for SN~2014J, 
and it has distinct substructures. 
Broad lines at shorter wavelengths, and dominated by [Co~III] line emission, do not show velocity shifts
between SNe~2011fe and 2014J. This fits nicely into the model of \citet{mae10b}, where low-ionisation lines are expected from 
asymmetrically distributed matter in the centre, whereas higher-ionisation lines originate further out where spherical symmetry is more 
likely. Both SNe~2011fe and 2014J have a slow decline rate of the velocity of the Si~II~$\lambda6355$ absorption 
trough just after $B-$band maximum. SN~2011fe fits well into the general picture that such supernovae have blueshifted nebular 
emission, while SN~2014J belong to a minority which instead have redshifted nebular emission. SN~2014J also has a larger velocity 
($\approx 11750 \kms$) of this trough at $B-$band maximum than usual.

Although we cannot count on being blessed with more very nearby SNe~Ia like SNe~2011fe and 2014J in the near future, late spectra
have now been obtained for more than just a handful SNe. As shown in Fig. \ref{fig:ablate}, useful constraints on progenitor systems
can be put on SNe~Ia also at $20-30$ Mpc, using our method. Concerted multi-wavelength efforts should be able to narrow
down possible progenitor systems of SNe~Ia. Both very early observations in order to constrain/detect possible interaction with
a binary or circumstellar gas, as well as late observations in the optical/infrared to constrain emission from ablated
gas and in radio to map circumstellar/interstellar gas, are needed.

\begin{acknowledgements}
We thank Jussi Harmanen, Tapio Pursimo and Ditte Slumstrup for making the spectroscopic observations at the NOT, and
Rahman Amanullah for discussions. 
PL acknowledges support from the Swedish Research Council. This research has made use of the NASA/IPAC Extragalactic 
Database (NED) which is operated by the Jet Propulsion Laboratory, California Institute of Technology, under contract with the 
National Aeronautics and Space Administration.
\end{acknowledgements}


\end{document}